\documentclass[preprint,showpacs,preprintnumbers,amsmath,amssymb]{revtex4}

\usepackage{graphicx}
\usepackage{epsfig}
\usepackage{bm}
\usepackage{amsfonts}
\usepackage{subfigure}
\usepackage{multirow}
\usepackage{array}
\usepackage{diagbox}
\usepackage{makecell}
\usepackage{slashbox}

\begin{document}

\title[]{Galileon intermediate inflation}

\author{Zeinab Teimoori$^{1}$\footnote{zteimoori16@gmail.com} and Kayoomars Karami$^{1,2}$\footnote{kkarami@uok.ac.ir}}
\address{$^{1}$\small{Department of Physics, University of Kurdistan, Pasdaran Street, P.O. Box 66177-15175, Sanandaj, Iran\\
$^2$Research Institute for Astronomy and Astrophysics of Maragha (RIAAM), P.O. Box 55134-441, Maragha, Iran}}

\begin{abstract}
We investigate intermediate inflation in the framework of a Galileon scalar field. To this aim, we first obtain the
inflationary observables, including the scalar spectral index, the tensor-to-scalar ratio, the running of the scalar
spectral index, as well as the non-Gaussianity parameters. Then, we examine the observational viability of the
intermediate inflation within the framework of a Galileon scenario. Our results show that although the prediction of
intermediate inflation in the standard framework is completely ruled out by the Planck 2015 observations, it can be
put inside the allowed regions of the Planck 2015 TT,TE,EE+lowP data in the Galileon setting. Moreover, we
determine the parameter space of the Galileon intermediate inflation for which the model is consistent with the
Planck 2015 data. Besides, we derive the consistency relation in the Galileon scenario, and find that it differs from
the standard inflation. We also estimate the running of the scalar spectral index and find that it is in well agreement
with the 95\% CL constraint of the Planck 2015 results. Finally, we evaluate the local, equilateral, orthogonal, and
enfolded non-Gaussianity parameters, and conclude that not only is the shape of non-Gaussianity approximately
close to the equilateral one, but that it also satisfies the 68\% CL  bound from the Planck 2015 data. 
\end{abstract}



\maketitle


\section{Introduction}\label{sec:int}
The inflationary paradigm, which was first proposed in the early 1980s, is a theoretical framework to describe the early stages of the evolution of the universe. Inflation addresses and partially solves some problems of the standard Big Bang theory such as the flatness problem, the horizon problem, and the magnetic monopole problem \cite{Starobinsky1980, Kazanas1980, Sato1981a, Sato1981, Guth1981, Linde1982, Albrecht1982, Linde1983}. Furthermore, inflation as an important part of modern cosmology provides the most convincing explanation for the origin of the anisotropy observed in the CMB radiation, as well as for the Large Scale Structure (LSS) formation in the universe \cite{Mukhanov1981, Hawking1982, Starobinsky1982, Guth1982}. According to the standard inflationary scenario, a canonical scalar field that is minimally coupled to the Einstein gravity, can explain how our universe expands at an accelerating rate during the inflationary era \cite{Liddle2000,Lemoine2000}. In the context of inflation, two different approaches have  usually been used to study inflationary scenarios: one based on scale factor, and the other case on the potential of scalar field inspired by particle physics. So far, different versions of inflation models with specific potentials or scale factors in the framework of standard inflationary scenario have been studied in the light of observational data \cite{Hossain2014,Martin2014a,Martin2014b,Geng2015,Huang2016}.

An interesting class of inflationary models, named $G$-inflation or Galileon models of inflation, have been discussed by many authors in which inflation is derived by the Galileon field  \cite{Kobayashi2010,Mizuno2010,Gao2011,Burrage2011,Kamada2011,Kobayashi2011,Felice2011,Kobayashi11,Felice011,Felice11,Ohashi2012,Tsujikawa2013,Renaux-Petel2013,Creminelli2011,Unnikrishnan2014,
Felice2013,Choudhury2015,Saitou2015,Kumar2016,Asadi2016,Bazrafshan2017,Motaharfar2017}. The Galileon field is a scalar field whose action is invariant under the Galilean symmetry $\partial_{\mu}\phi\rightarrow \partial_{\mu}\phi+ b_{\mu}$ in the Minkowski spacetime \cite{Nicolis2009,Deffayet2009,Deffayet09}. This scalar field theory was first proposed in \cite{Nicolis2009}, and was inspired by the Dvali-Gabadadze-Porrati (DGP) model. The DGP scenario is a braneworld model that is plagued by ghost-like instabilities \cite{Dvali2000,Deffayet2002,Koyama2005}. The authors of \cite{Nicolis2009} showed that in 4D, there exist only five Lagrangian terms which admit this type of symmetry in the Minkowski spacetime. The analysis used in \cite{Nicolis2009} is extended to the curved spacetime in \cite{Deffayet2009,Deffayet09} and the authors showed that the Galileon invariance of the original model is broken by the resulting theory.
Deffayet \textit{et al.} \cite{Deffayet2011} derived the most general action in those theories in which the scalar field $\phi$ is non-minimally coupled to the metric. The remarkable property of the Galileon Lagrangian is that it yields second-order gravitational and scalar field equations. This property is important because higher-derivative theories contain additional degrees of freedom and usually suffer from significant problems, such as negative energies and related instabilities \cite{Ostrogradski1850}. For more  about the Galileon scalar field theory, see, e.g.  \cite{Chow2009,Silva2009,Kobayashi10,Kobayashi010,Gannouji2010,Felice2010,Felice10,Felice010,Gao2015,Carrillo2016,Saltas17,
Saltas2017,Deffayet2017,Banerjee2017,Banerjee17,Kolevatov2017}.

In \cite{Kobayashi11,Charmousis2012}, it was shown that the Galileon action derived by Deffayet \textit{et al.} \cite{Deffayet2011} is equivalent to that derived by Horndeski in 1974 \cite{Horndeski1974}. Nowadays, the Horndeski theory is known as the most general scalar-tensor theory with second-order field equations, and it covers a wide range of gravitational theories, such as standard canonical inflation \cite{Linde1982,Linde1983}, non-minimally coupled models \cite{Fakir1990,Bezrukov2008}, non-canonical scalar field models \cite{Nakayama2010,Felice2011}, Galileon inflation \cite{Kobayashi2010}, Brans-Dicke theory \cite{Brans1961}, field derivative couplings to gravity \cite{Amendola1993} and $k$-inflation \cite{Armendariz1999,Garriga1999}.

In this work, we focus on the $G$-inflation with the Galileon term $G(\phi,X)\Box \phi\propto X^{n} \Box \phi$ \cite{Felice2011,Felice011} and examine the viability of intermediate inflation in light
of the Planck 2015 data \cite{Planck2015}. This power-law Galileon interaction corresponds to the generalization of the original
 Galileon field theory, $ X \Box \phi$, which was first proposed in the Minkowski spacetime \cite{Nicolis2009} and extended later to the curved spacetime by Deffayet \textit{et al.} \cite{Deffayet2009,Deffayet09}. For the case $n=1$, the coupling $X \Box \phi$ has extensively  studied in the literature \cite{Kobayashi2010,Mizuno2010,Kamada2011,Kobayashi2011,Kobayashi11,Felice11,Ohashi2012}. Intermediate inflation is described with a scale factor in the form of $a(t)\propto \exp [A({M_{\rm pl}}\,t)^\lambda]$ where $A>0$ and $0<\lambda<1$. Note that in the standard scenario, intermediate inflation is driven by an inverse power-law potential  \cite{Barrow90,Barrow-Saich90,Muslimov1990,Barr-Lidd1993,Vallinotto2004,Rendall2005,Barrow2006,Campo14} and it is not consistent with the Planck 2015 data \cite{Rezazadeh2015,Rezazadeh2016,Nazavari2016,Teimoori2017,Amani2018}. This motivates us to investigate whether the intermediate inflation in the Galileon setting can be resurrected in light of the Planck 2015 results.

The present paper is structured as follows.
In Sec. \ref{review:sec}, we briefly review the Galileon inflationary scenario and obtain all the necessary equations describing the mechanism of inflation. In Secs. \ref{sec:int}, we apply the results of Sec. \ref{review:sec} for the intermediate inflation and investigate the viability of this model with respect to the Planck 2015 data. Section \ref{sec:con} is devoted to our concluding remarks.
\section{Inflation with the Galileon scalar field}\label{review:sec}

The Galileon inflation model is characterized by the following general action \cite{Kobayashi2010,Deffayet2011}
\begin{equation}\label{action}
S= \int d^{4}x \sqrt{-g}\left[\frac{1}{2} R + {\cal L}_{\phi} \right],
\end{equation}
where
\begin{equation}\label{Lagrangian}
{\cal L}_{\phi}\equiv K(\phi, X)-G(\phi, X)\Box \phi,
\end{equation}
is the scalar field Lagrangian, $g$ is the determinant of the metric $ g_{{\mu}{\nu}}$ and $R$ is the Ricci scalar.
Throughout this paper, we work in units in which the reduced Planck mass is set equal to unity, i.e. $M_{\rm pl}=(8\pi G)^{-1/2}=1$.
Here, $K(\phi,X)$ and $G(\phi,X)$ are general functions of $\phi$ and the kinetic term $X\equiv -\frac{1}{2}g^{\mu\nu}\phi_{,\mu}\phi_{,\nu}$.
Note that the Galileon scalar filed Lagrangian (\ref{Lagrangian}) is nothing but the kinetic gravity braiding \cite{Deffayet2010}. In \cite{Ezquiaga2017}, it was shown that the kinetic gravity braiding is compatible with the recent observation of GW170817 detected by the LIGO-VIRGO Collaboration \cite{Virgo-Ligo2017}. In \cite{Deffayet2010}, the authors showed that for any form of functions $K$ and $G$, Lagrangian (\ref{Lagrangian}) leads to the second-order field equations.

In the following, we review briefly the basic relations governing the theory of cosmological perturbation in the Galileon scenario. To this aim, we follow the approach of \cite{Tsujikawa2013} in which we set $P(\phi, X)=K(\phi, X), G_3(\phi, X)=G(\phi, X), G_4=0=G_5$ to transform the results of \cite{Tsujikawa2013} to our model. The power spectrum of the primordial scalar perturbation ${\cal P}_{s}$ in the slow-roll limit reads \cite{Tsujikawa2013}
\begin{equation}\label{Ps}
{\cal P}_{s}\simeq\frac{H^2}{8 \pi ^{2}c_{s}^3\,q_{s}}\Big|_{c_{s}k=aH}\,,
\end{equation}
where ${\cal P}_{s}$ is evaluated at the
 time of sound horizon exit, i.e. $c_{s}k=aH$ in which $k$ is a comoving wavenumber. The recent value of the scalar perturbation amplitude at the horizon crossing has been reported by the Planck collaboration as $\ln\left[10^{10}{\cal P}_{s}\left(k_{*}\right)\right]=3.094\pm0.034$ (Planck 2015 TT,TE,EE+lowP data) \cite{Planck2015}, where $k_{*}=0.05\,{\rm Mpc}^{{\rm -1}}$ is the pivot scale.

In Eq. (\ref{Ps}), the parameter $q_s$ and the scalar propagation speed squared $c_{s}^{2}$ are given by
 \begin{equation}\label{qs}
 q_{s}= \delta_{X}+2 \delta_{XX}+6 \delta_{GX}+6 \delta_{GXX}-2 \delta_{G\phi},
 \end{equation}
\begin{equation}\label{cs}
c_{s}^{2}=\frac{\varepsilon_{s}}{q_{s}},
\end{equation}
where
\begin{equation}\label{SRP}
\delta_{X}\equiv \frac{K_{,X} X}{H^2},~~~\delta_{XX}\equiv \frac{ K_{,XX} X^2}{H^2},~~~ \delta_{GX}\equiv \frac{G_{,X}\dot{\phi} X}{ H},~~~
 \delta_{GXX}\equiv \frac{G_{,XX}\dot{\phi} X^2}{H},~~~\delta_{G\phi}\equiv \frac{G_{,\phi} X}{H^2},
\end{equation}
are the slow-roll parameters in the Galileon scenario and
\begin{equation}\label{epsilon:s1}
\varepsilon_{s}\equiv\delta_{X}+ 4\delta_{GX}-2\delta_{G\phi}.
 \end{equation}
In the above relations, $X=\dot{\phi}^2/2$ and a dot indicates derivative with respect to the cosmic time, $t$.
We also use the notations $({,\phi})$ and $({,X})$ to denote ${\partial }/{\partial \phi}$ and ${\partial }/{\partial X}$, respectively.

The scale-dependence of the scalar perturbation spectrum is determined by the scalar spectral index $n_{s}$ which in the Galileon scenario it is given by  \cite{Tsujikawa2013}
\begin{equation}\label{ns1}
n_{s}-1\equiv \frac{d\ln{{\cal P}_{s}}}{d\ln{k}}\Big|_{c_{s}k=aH}\simeq\frac{\dot{{\cal P}_s}}{H {\cal P}_{s}}\simeq {-2 \varepsilon_{1}}-\eta_{s1}-s_{1},
\end{equation}
where
 \begin{equation}\label{epsilon}
 \varepsilon_{1}\equiv -\frac{\dot H}{H^2}\simeq\delta_{X}+3\delta_{GX}-2\delta_{G\phi},
\end{equation}
and
\begin{equation}\label{eta:s}
\eta_{s1} \equiv \frac{\dot \varepsilon_{s}}{H\varepsilon_{s}},~~~s_{1}\equiv \frac{\dot{c_{s}}}{H c_{s}}.
\end{equation}
 Observational value of the scalar spectral index is $ n_{s}=0.9644 \pm 0.0049 $ (68\% CL, Planck 2015 TT,TE,EE+lowP data) \cite{Planck2015}. Note that in the first approximation in Eq. (\ref{ns1}), we have used the approximation $d\ln k \thickapprox H dt$. This is because of during slow-roll inflation, the Hubble parameter $H$ and the sound speed $c_s$  vary much slower than the scale factor $a$ of the universe \cite{Garriga1999}. Therefore, using the relation $c_s k = a H$ which is valid around the sound horizon exit, we can write $d\ln k \thickapprox H dt$.

The tensor power spectrum in the Galileon inflationary setting is given by \cite{Tsujikawa2013}
 \begin{equation}\label{Pt}
{\cal P}_{t}\simeq\frac{2 H^2}{\pi ^{2}}\Big|_{k=aH}\,,
\end{equation}
 which is evaluated at the time of horizon crossing specified by $k = aH$. This time is not exactly the same as the time of sound horizon
crossing for which $c_sk = aH$, but to lowest order in the slow-roll
parameters the difference is negligible \cite{Garriga1999}.

The tensor spectral index, $n_{t}$, which determines the scale-dependence of the tensor power spectrum in the Galileon inflation takes the form  \cite{Tsujikawa2013}
 \begin{equation}\label{nt1}
n_{t}\equiv \frac{d\ln{{\cal P}_{t}}}{d\ln{k}}\Big|_{k=aH}\simeq -2 \varepsilon_{1}.
\end{equation}
In the Galileon scenario, using Eqs. (\ref{Ps}), (\ref{cs}), and (\ref{Pt}), we can find the tensor-to-scalar ratio as
\begin{equation}\label{r:epsilon}
r \equiv\frac{{\cal P}_{t}}{{\cal P}_{s}}\simeq 16 c_{s}\,\varepsilon_{s}\,.
\end{equation}
The Planck 2015 data sets an upper bound on the tensor-to-scalar ratio as $r<0.149$ (95\% CL, Planck 2015 TT,TE,EE+lowP data) \cite{Planck2015}. Using Eqs. (\ref{epsilon:s1}), (\ref{epsilon}), (\ref{nt1}), and (\ref{r:epsilon}), the consistency relation between $r$ and $n_{t}$ in the Galileon inflation is obtained as
\begin{equation}\label{consistency}
r=-8 c_{s}(n_{t}-2\delta_{GX}),
\end{equation}
which is different from the consistency relation $r=-8n_t$ in the standard inflation. Basically, the consistency relation can be used as a powerful tool to test different inflationary scenarios \cite{Lidsey1997}. However, to this aim, a measurement of $n_t$ is required; if such information becomes available, it will probably easier to distinguish between different inflationary models via the $r-n_t$ plane. However, it is worth noting that unfortunately $n_t$ will be so difficult to measure, so we have to wait for future observations. If the standard consistency relation, $r=-8\,n_t$, could be ruled out, then it would imply a non-standard inflation model or a multiple field inflation model.

Note that in order to have a Galileon model free of ghosts and Laplacian instabilities, we require the following conditions, respectively  \cite{Tsujikawa2013}
\begin{equation}\label{ghost}
q_{s}>0,~~~~~~~c_{s}^2>0.
\end{equation}

Another significant observable predicted by inflation, is the non-Gaussianity parameter which can be used as a strong tool to distinguish between various inflationary models \cite{Babich2004}. There are several different non-Gaussianities containing the local, equilateral, orthogonal, and enfolded shapes, that depend on the wave numbers $k_1$, $k_2$ and $k_3$, with the momentum triangle
satisfying $k_1+k_2+k_3=0$. Note that the local and orthogonal shapes, respectively, arise in the multi-fields inflation \cite{{Bartolo2002,Sasaki2008}} and in the models with non-standard initial states \cite{Chen2007, Holman2008}. The equilateral non-Gaussianity is generated by the single field inflationary models with non-canonical kinetic term like the Galileon scalar field in our model \cite{Alishahiha2004, Chen2007,Li2008,Tolley2010}. The local, equilateral, orthogonal and enfolded non-Gaussianity parameters are given by \cite{Felice2013}
 \begin{equation}\label{local}
 f_{\rm NL}^{\rm local}\Big|_{k_1=k_2,~k_3\rightarrow 0}=\frac{5}{12}(1-n_s),
 \end{equation}
\begin{equation}
\label{fNL1}
 f_{\rm NL}^{\rm equil}\Big|_{k_1=k_2=k_3}=
  \frac{85}{324}\left(1-\frac{1}{c_{s}^2}\right)-\frac{10}{81}\,\frac{\mu}{\Sigma}+\frac{20}{81\varepsilon_{s}}
  \big(\delta_{GX}+\delta_{GXX}\big)
  +\frac{65}{162\,c_{s}^2\, \varepsilon_{s}}\,\delta_{GX},
  \end{equation}
 \begin{equation}
\label{fNL-ortho1}
 f_{\rm NL}^{\rm ortho}\Big|_{k_1=2k_2=2k_3} =
  \frac{259}{1296}\left(1-\frac{1}{c_{s}^2}\right)+\frac{1}{648}\,\frac{\mu}{\Sigma}-\frac{1}{324\varepsilon_{s}}
  \big(\delta_{GX}+\delta_{GXX}\big)
  +\frac{65}{162\,c_{s}^2\, \varepsilon_{s}}\,\delta_{GX},
 \end{equation}
\begin{equation}
\label{fNL-enfold1}
 f_{\rm NL}^{\rm enfold}\Big|_{k_1=k_2+k_3} =
  \frac{1}{32}\left(1-\frac{1}{c_{s}^2}\right)-\frac{1}{16}\,\frac{\mu}{\Sigma}+\frac{1}{8\varepsilon_{s}}
  \big(\delta_{GX}+\delta_{GXX}\big),
 \end{equation}
where the quantities $\mu$ and $\Sigma$ are given by Eqs. (3.11) and (3.12) in \cite{Felice2013} with settings $P(\phi, X)=K(\phi, X), G_3(\phi, X)=G(\phi, X), G_4=0=G_5$.
Note that the enfolded shape (\ref{fNL-enfold1}) can also be expressed in terms of the equilateral and orthogonal ones as follows \cite{Creminelli2011,Senatore2010}
 \begin{equation}\label{enfold0}
f_{\rm NL}^{\rm enfold}=\frac{1}{2}(f_{\rm NL}^{\rm equil}- f_{\rm NL}^{\rm ortho}).
\end{equation}
The Planck Collaboration \cite{Planck2015-nonGauss} found $f_{\rm NL}^{\rm local} = 0.8 \pm 5.0$, $f_{{\rm{NL}}}^{{\rm{equil}}} = - 4 \pm 43$, $f_{{\rm{NL}}}^{{\rm{ortho}}} =- 26 \pm 21$, $f_{{\rm{NL}}}^{{\rm{enfold}}} = 11\pm 32$ from both temperature and polarization data at 68\% CL.

So far, we study the basic relations governing the general Galileon inflationary scenarios. Hereafter, we focus on a particular case and investigate a concrete model.
Let us consider the potential-driven inflation as \cite{Felice011,Felice2011,Ohashi2012,Tsujikawa2013}
\begin{equation}\label{K-form}
K(\phi, X)=X-V(\phi)\,,
\end{equation}
where $V(\phi)$ is the scalar field potential.
Also from \cite{Felice2011,Felice011}, we take the functional form of $G(\phi, X)$ as
\begin{equation}\label{G-form}
G(X)=\frac{X^n}{M^{4n-1}}\,,
\end{equation}
where $n>0$ is an integer constant. Also, $M>0$ is constant and has a dimension of mass. Note that the case $n=1$ has been studied in \cite{Ohashi2012,Tsujikawa2013}.

For the $G$ function (\ref{G-form}), from the fifth relation in Eq. (\ref{SRP}) we have $ \delta_{G\phi}=0$. Then, the slow-roll parameter $\varepsilon_{1}$ in Eq. (\ref{epsilon})  can be written as
\begin{equation}\label{epsilon:A}
\varepsilon_{1}\simeq \left(1+\cal D\right)\delta_{X},
\end{equation}
where
\begin{equation}\label{A}
{\cal D}\equiv 3\,\frac{\delta_{GX}}{\delta_{X}}=\frac{3 n}{2^{n-1} M^{4n-1}}\,\dot{\phi}^{2n-1}H,
\end{equation}
and we have used $X=\dot{\phi}^2/2$ in the second equality.

Under the slow-roll approximation, the first Friedmann equation (4) in \cite{Tsujikawa2013}, reduces to
\begin{equation}\label{FR:SR}
3 H^2\simeq V(\phi),
\end{equation}
which is the same as that obtained in the standard scenario. Taking the time derivative of Eq. (\ref{FR:SR}), also using Eq. (\ref{epsilon:A}), the definition $\varepsilon_{1}$ in Eq. (\ref{epsilon}) and the first relation in Eq. (\ref{SRP}), it follows that
\begin{equation}\label{Field:SR}
3 H \dot{\phi}(1+{\cal D})+V_{,\phi}\simeq 0.
\end{equation}
From the third and fourth relations in Eq. (\ref{SRP}) and using Eq. (\ref{G-form}), it follows that $\delta_{GXX}=(n-1)\,\delta_{GX}$. Also, with the help of Eq. (\ref{K-form}) and the second  relation in Eq. (\ref{SRP}), we find  $\delta_{XX}=0$, and consequently the parameter $q_{s}$ given by Eq. (\ref{qs}) reduces to
\begin{align}\label{qs2}
q_{s}=\delta_{X}+6 n \delta_{GX}=\delta_{X}(1+2 n {\cal D}).
\end{align}
Inserting Eqs. (\ref{epsilon:s1}) and (\ref{qs2}) into (\ref{cs}), it follows that
\begin{align}\label{cs2}
c_{s}^2=\frac{\delta_{X}+4 \delta_{GX}}{\delta_{X}+6 n \delta_{GX}}=\frac{1+\frac{4}{3}{\cal D}}{1+2 n {\cal D}}.
\end{align}
Note that from the first relation in Eq. (\ref{SRP}), and using Eq. (\ref{K-form}) and  $X=\dot{\phi}^2/2$, we have $\delta_{X}=\frac{\dot{\phi}^{2}}{{2 H^2}}>0$. Therefore, to avoid the ghosts and
Laplacian instabilities in our model, from Eqs. (\ref{qs2}) and (\ref{cs2}) the conditions (\ref{ghost}) are preserved, provided that ${\cal D}>0$, keeping in our mind that $n>0$ has an integer value. Consequently, from Eq. (\ref{A}), the condition ${\cal D}>0$ requires that $\dot{\phi}>0$, therefore, Eq. (\ref{Field:SR}) leads to $V_{,\phi}<0$.

Using Eqs. (\ref{FR:SR}) and (\ref{Field:SR}), one can find the slow-roll parameter $\delta_{X}$ in Eq. (\ref{SRP}) as
\begin{equation}\label{deita:x}
\delta_{X}\simeq \frac{\varepsilon_{\rm v}}{(1+{\cal D})^2},
\end{equation}
where
\begin{equation}\label{epsilon:v}
\varepsilon_{\rm v}\equiv\frac{1}{2}\left(\frac{V_{,\phi}}{V}\right)^2.
\end{equation}
Using Eqs. (\ref{epsilon:A}) and (\ref{deita:x}), we have
\begin{equation}\label{epsilon:v2}
\varepsilon_{\rm v}\simeq \varepsilon_{1}(1+{\cal D}).
\end{equation}
Applying Eqs. (\ref{FR:SR}) and (\ref{qs2}), (\ref{cs2}), (\ref{deita:x}) and (\ref{epsilon:v}), the power spectrum (\ref{Ps}) can be obtained as
\begin{equation}\label{Ps2}
{\cal P}_{s}=\frac{1}{12\,\pi^2}\left(\frac{V^3}{V_{,\phi}^2}\right)\frac{(1+{\cal D})^2\,(1+2 n {\cal D})^{1/2}}{\Big(1+\frac{4}{3}{\cal D}\Big)^{3/2}}.
\end{equation}
From $n_{s}-1\simeq\dot{{\cal P}_s}/(H {\cal P}_{s})$, and then by using Eqs. (\ref{FR:SR}), (\ref{Field:SR}), (\ref{epsilon:v}) and (\ref{Ps2}) we can compute the scalar spectral index $n_{s}$ in our Galileon scenario as
\begin{equation}\label{ns3}
n_{s}-1\simeq\frac{1}{1+{\cal D}}\left(2 \eta_{\rm v}-6\varepsilon_{\rm v}\right)+\frac{\dot{{\cal D}}}{H}\left(\frac{2}{1+{\cal D}}+\frac{n}{1+2 n {\cal D}}-\frac{2}{1+\frac{4}{3}{\cal D}}\right),
\end{equation}
where
\begin{equation}\label{eta:v}
\eta_{\rm v}\equiv\frac{V_{,\phi \phi}}{V}.
\end{equation}
Considering the definition (\ref{A}) and using Eqs. (\ref{epsilon:s1}) and (\ref{deita:x}), it is easy to get
\begin{equation}\label{epsilon:s2}
\varepsilon_{s}=\varepsilon_{\rm v}\,\frac{1+\frac{4}{3}{\cal D}}{(1+{\cal D})^2}.
\end{equation}
From Eqs. (\ref{cs2}) and (\ref{epsilon:s2}), the tensor-to-scalar ratio (\ref{r:epsilon}) yields
\begin{equation}\label{r3}
r=16\,\varepsilon_{\rm v}\,\frac{(1+\frac{4}{3}{\cal D})^{3/2}}{(1+2 n {\cal D})^{1/2}(1+{\cal D})^2}.
\end{equation}

For the Galileon model described by Eqs. (\ref{K-form}) and (\ref{G-form}), since $K_{,XX}=K_{,XXX}=G_{,\phi X}=G_{,\phi X X}=0$, from the third and fourth relations in Eq. (\ref{SRP}), the quantity $\mu$  which is given by Eq. (3.12) in \cite{Felice2013}, can be written in the following form:
\begin{equation}\label{mu2}
 \mu=H^2\big(\delta_{GX}+5\delta_{GXX}+2\delta_{GXXX}\big),
\end{equation}
where
\begin{equation}\label{GXXX}
\delta_{GXXX}\equiv\frac{G_{,XXX}\dot{\phi}X^3}{H}.
\end{equation}
Using the second and fifth relations in Eq. (\ref{SRP}), it is easy to show that $\delta_{G\phi}=\delta_{XX}=0$; then, at first order in slow-roll parameters, one can find $\Sigma$ given by  Eq. (3.11) in \cite{Felice2013}, as follows:
\begin{equation}\label{sigma2}
\Sigma=H^2\big(\delta_{X}+6\delta_{GX}+6\delta_{GXX}\big).
\end{equation}
From Eq. (\ref{GXXX}) and using the third relation in Eq. (\ref{SRP}), we find $\delta_{GXXX}=(n-1)(n-2)\delta_{GX}$. Then, with the help of Eqs. (\ref{epsilon:s1}), (\ref{cs2}), (\ref{mu2}) and (\ref{sigma2}), the definition (\ref{A}) and the relation $\delta_{GXX}=(n-1)\delta_{GX}$, the non-Gaussianities  (\ref{fNL1}), (\ref{fNL-ortho1}) and (\ref{fNL-enfold1}) reduce to the following forms, respectively,
\begin{align}
\label{fNL2}
 f_{\rm NL}^{\rm equil} =\frac{85}{162}\frac{(2-3n){\cal D}}{3+4{\cal D}}+\frac{{\cal D}}{243}\left(\frac{10 n(1-2 n)}{1+2 n{\cal D}}+\frac{60n}{3+4{\cal D}}+ \frac{585(1+2 n {\cal D})}{2(3+4{\cal D})^2}\right),\\
 \label{fNL-ortho2}
 f_{\rm NL}^{\rm ortho} =\frac{259}{648}\frac{(2-3n){\cal D}}{3+4{\cal D}}-\frac{{\cal D}}{486}\left(\frac{ n(1-2 n)}{4(1+2 n{\cal D})}+\frac{3n}{2(3+4{\cal D})}- \frac{585(1+2 n {\cal D})}{(3+4{\cal D})^2}\right),\\
 \label{fNL-enfold2}
 f_{\rm NL}^{\rm enfold} =\frac{1}{16}\frac{(2-3n){\cal D}}{3+4{\cal D}}+\frac{{\cal D}}{24}\left(\frac{ n(1-2 n)}{2(1+2 n{\cal D})}+\frac{3n}{3+4{\cal D}}\right).
 \end{align}
Also, the local non-Gaussianity $f_{\rm NL}^{\rm local}$ is obtained by replacing Eq. (\ref{ns3}) into (\ref{local}). From the definition (\ref{A}), and using Eqs. (\ref{nt1}) and (\ref{epsilon:A}), one can obtain
\begin{equation}\label{deltaGX}
\delta_{GX}=-\frac{1}{6}\left(\frac{{\cal D}}{1+{\cal D}}\right)n_t.
\end{equation}
Substituting Eq. (\ref{deltaGX}) into (\ref{consistency}), the consistency relation reads
\begin{equation}\label{consistency2}
r=-8\, c_{s}\,n_{t}\left(1+\frac{\cal{D}}{3(1+\cal{D})}\right).
\end{equation}
Note that for the case $n=1$, Eqs. (\ref{Ps2}), (\ref{ns3}) and (\ref{r3}) are transformed to the corresponding results obtained in \cite{Ohashi2012}.

It is also worth to mentioning that in the limit of ${\cal D}\,\ll 1$ (or $M\rightarrow \infty$), i.e., $\delta_{X} \gg |\delta_{GX}|$,
 Eqs. (\ref{cs2}), (\ref{Ps2}), (\ref{ns3}), (\ref{r3}), and (\ref{consistency2}) reduce to the well-known relations in the standard inflation. This is to be expected, because in the regime ${\cal D}\,\ll 1$, the standard kinetic term $X$ appearing in the scalar field Lagrangian (\ref{Lagrangian}) dominates over the Galileon self-interaction term $G(\phi,X) \,\Box \phi\propto  X^{n} \,\Box \phi$. Therefore, our Galileon inflationary model described by Eqs. (\ref{K-form}) and (\ref{G-form}) in the limit ${\cal D}\,\ll 1$ recovers the results of standard inflation.

\subsection{Galileon inflation in the regime ${\cal D}\gg 1$ (or $M\rightarrow 0$) }

Now, we turn to investigate our model in the regime, where the Galileon term $G(\phi,X)\Box \phi\propto  X^{n} \Box \phi$ in the scalar field Lagrangian (\ref{Lagrangian}), dominates over the standard kinetic term $X$. In this case, one can get $|\delta_{GX}| \gg \delta_{X}$, which yields ${\cal D}\gg 1$ (or $M\rightarrow 0$).
In this limit, Eq. (\ref{Field:SR}) takes the form $3\,H\dot{\phi}\,{\cal D}+V_{,\phi}\simeq0$. Now, with replacing ${\cal D}$ from Eq. ({\ref{A}}), one can solve the resulting equation for $\dot{\phi}$ which gives
\begin{equation}\label{phi:dot1}
\dot{\phi}=\pm\left(-\frac{M^{4n-1} 2^{n-1}}{9n H^2}~V_{,\phi}\right)^{\frac{1}{2n}}.
\end{equation}
Because  we require that $ \dot{\phi}> 0$ to avoid the ghosts and Laplacian instabilities, we choose the positive sign in Eq. (\ref{phi:dot1}).
Then, inserting it into (\ref{A}), we get
\begin{equation}\label{A1}
{\cal D}=\frac{1}{3}\left(-\frac{M^{4n-1} 2^{n-1}}{9 n}~\frac{H^{2n-2}}{{V_{,\phi}}^{2n-1}}\right)^{-\frac{1}{2n}}.
\end{equation}
Taking the time derivative of Eq. (\ref{A1}) and using the definition $\varepsilon_{1}$ in Eq. (\ref{epsilon}), and also Eqs. (\ref{FR:SR}), (\ref{epsilon:v2}), (\ref{eta:v}) and the relation $3\,H\dot{\phi}\,{\cal D}+V_{,\phi}\simeq0$, we find
\begin{equation}\label{1}
\frac{\dot{{\cal D}}}{H}=\left(1-\frac{1}{n}\right)\varepsilon_{\rm v}-\left(1-\frac{1}{2n}\right)\eta_{\rm v}.
\end{equation}
Finally, in the regime ${\cal D}\,\gg 1$, the power spectrum of scalar perturbations (\ref{Ps2}), the scalar spectral index (\ref{ns3}), the tensor-to-scalar ratio (\ref{r3}), the non-Gaussianity parameters (\ref{fNL2}), (\ref{fNL-ortho2}), (\ref{fNL-enfold2}), and the consistency relation (\ref{consistency2}) take the following forms:
\begin{align}
\label{Ps:A}
&{\cal P}_{s} =\left(\frac{1}{16\,\pi^2}~\sqrt{\frac{3\,n}{2}}~\right)\left(\frac{V^3}{V_{,\phi}^2}\right)\,{\cal D}, \\
\label{ns:A}
   n_{s}-1 =&\left(1+\frac{1}{2 n}\right)\frac{\eta_{\rm v}}{\cal D}-\left(5+\frac{1}{n}\right)\frac{\varepsilon_{\rm v}}{\cal D}, \\
   \label{r:A}
& r = \left(\frac{64}{3}~\sqrt{\frac{2}{3n}}~\right)\frac{\varepsilon_{\rm v}}{\cal D},\\
 \label{fNL:A}
 & f_{\rm NL}^{\rm equil} = \left(\frac{275}{972}\right)-\left(\frac{865}{3888}\right)n,\\
 \label{fNLortho:A}
 & f_{\rm NL}^{\rm ortho} = \left(\frac{97}{486}\right)-\left(\frac{1163}{7776}\right)n,\\
 \label{fNLenfold:A}
 & f_{\rm NL}^{\rm enfold} = \left(\frac{1}{24}\right)-\left(\frac{7}{192}\right)n,\\
 \label{consistency:A}
& r=-\left(\frac{32}{3}\right) c_s n_t,
\end{align}
where we have used Eq. (\ref{1}) in deriving Eq. (\ref{ns:A}). Note that for $n\gg1$, Eq. (\ref{fNL:A}) reduces to $f_{\rm NL}^{\rm equil} = -\frac{865}{3888}\,n$, which is same as the result obtained in \cite{Felice2011}. Besides, in the limit ${\cal D}\,\gg 1$, Eq. (\ref{cs2}) reduces to
\begin{equation}\label{cs2:A}
c_{s}^2=\frac{2}{3 n},
\end{equation}
which shows that $c_s^2$ is positive. Note that the condition $c_s^2>0$ is needed to avoid  Laplacian instability of scalar perturbations in our model.

\section{Intermediate inflation driven by a Galileon scalar field}\label{sec:int}

In \cite{Rezazadeh2015,Rezazadeh2016,Nazavari2016,Teimoori2017,Amani2018}, it was shown that in the framework of standard inflationary scenario, the result of intermediate inflation is completely ruled out by the Planck 2015 TT,TE,EE+lowP data \cite{Planck2015}. This motivates us to investigate intermediate inflation in the Galileon scenario to see whether it can be resurrected in light of the Planck 2015 results.

Intermediate inflation is characterized by the following scale factor \cite{Barrow90,Barrow-Saich90,Muslimov1990,Barr-Lidd1993,Vallinotto2004,Rendall2005,Barrow2006,Campo14}:
\begin{equation}\label{scale}
a(t)= a_{0} \exp\left(A\,t^\lambda\right),
\end{equation}
where $A>0$ and $0<\lambda<1$. The potential driving the intermediate inflation in the standard inflation model has an inverse power-law form as $V(\phi)\propto \phi^{-p}$, where $p=4(1-\lambda)/\lambda$.
 For the scale factor (\ref{scale}), the Hubble parameter, $H$, takes the form
\begin{equation}\label{Hubble}
H(t)= \frac{A \,\lambda}{t^{1-\lambda}}.
\end{equation}
Inserting Eq. (\ref{Hubble}) into (\ref{A}), we get
\begin{equation}\label{A2}
{\cal D}=\left(\frac{6  A n \lambda}{2^n M^{4n-1}}\right)\left(\frac{\dot{\phi}^{2n-1}}{t^{1-\lambda}}\right).
\end{equation}
Using Eqs. (\ref{FR:SR}) and (\ref{Hubble}), the power spectrum (\ref{Ps2}) takes the form
\begin{equation}\label{Ps-t}
{\cal P}_{s}=\frac{1}{16\pi^2}\Big(\frac{A\,\lambda}{\lambda-1}\Big)^2\frac{(1+{\cal D})^2\,(1+2 n {\cal D})^{1/2}}{\Big(1+\frac{4}{3}{\cal D}\Big)^{3/2}}~t^{2\lambda}\dot{\phi}^2.
\end{equation}
Also, the slow-roll parameters (\ref{epsilon:v}) and (\ref{eta:v}) read
\begin{equation}\label{epsilon:eta}
\varepsilon_{\rm v}=\frac{(\lambda-1)^2}{t^2\dot{\phi}^2},\hspace{1.2cm}\eta_{\rm v}=\frac{2\left(1-\lambda\right)
\left[(3-2\lambda)\dot{\phi}+t\ddot{\phi}\right]}{t^2\dot{\phi}^3}.
\end{equation}
With the help of Eqs. (\ref{Hubble}), (\ref{A2}) and (\ref{epsilon:eta}), one can rewrite the scalar spectral index (\ref{ns3}) and  the tensor-to-scalar ratio (\ref{r3}) as
\begin{align}\label{ns-t}
n_{s}-1=\frac{4(1-\lambda)(\lambda \dot{\phi}+t \ddot{\phi})}{1+{\cal D}}~t^{-2}\dot{\phi}^{-3}
+\left(\frac{(2n-1)\ddot{\phi}\,t+
(\lambda-1)\dot{\phi}}{A\,\lambda}\right)\nonumber\\
\times\left(\frac{2{\cal D}}{1+{\cal D}}+\frac{n{\cal D}}{1+2 n {\cal D}}-\frac{2{\cal D}}{1+\frac{4}{3}{\cal D}}\right)~t^{-\lambda}\dot{\phi}^{-1},
\end{align}
and
\begin{align}\label{r-t}
r=\frac{16\,(\lambda-1)^2(1+\frac{4}{3}\,{\cal D})^{{3/2}}}{(1+{\cal D})^2(1+2\,n\,{\cal D})^{{1/2}}}~t^{-2}\dot{\phi}^{-2}.
\end{align}
Using the approximation $d\ln k \thickapprox H dt$ and Eq. (\ref{Hubble}); from Eq. (\ref{ns-t}), one can obtain the running of the scalar spectral index as
\begin{eqnarray}\label{dns-t}
\frac{d n_{s}}{d\ln k}=\frac{J_{3}}{1+{\cal D}}&+&J_{2}{\cal D}^2\left[\frac{J_1}{{\cal D}(1+{\cal D})^2}-2\,J_{2\,}\left(\frac{1}{(1+{\cal D})^2}+\frac{n^2}{(1+2\,n\,{\cal D})^2}-\frac{4}{3\Big(1+\frac{4}{3}\,{\cal D}\Big)^2}\right)\right]\nonumber\\&+&(J_4+J_{2}^2)\left(\frac{2}{1+{\cal D}}+\frac{n}{1+2\,n\,{\cal D}}-\frac{2}{1+\frac{4}{3}\,{\cal D}}\right),
\end{eqnarray}
where
\begin{align}
\label{J1}
 J_{1} \equiv   \frac{4(1-\lambda)(\lambda \dot{\phi}+t \ddot{\phi})}{t^{2}\dot{\phi}^{3}}, \\
  \label{J2}
  J_{2} \equiv \frac{t^{1-\lambda}}{A\,\lambda}\left[(2n-1)\left(\frac{\ddot{\phi}}{\dot{\phi}}\right)+\frac{\lambda-1}{t}\right],
\end{align}
\begin{align}\label{J3}
  J_{3}\equiv4 A \lambda (\lambda-1)\left[\frac{2\lambda}{\dot{\phi}^2}\,t^{\lambda-4}
  +\left(3\frac{\ddot{\phi}}{\dot{\phi}^4}-\frac{\dddot{\phi}}{\dot{\phi}^3}\right)t^{\lambda-2}+
  (1+2\lambda)\frac{\ddot{\phi}}{\dot{\phi}^3}~t^{\lambda-3}\right], \\
  \label{J4}
  J_{4}  \equiv\frac{\lambda(1-\lambda)}{t^2}+(1-2 n)\left[\left(\frac{\ddot{\phi}}{\dot{\phi}}\right)^2+(\lambda-1)\frac{\ddot{\phi}}{t \dot{\phi}}-\frac{\dddot{\phi}}{\dot{\phi}}\right].
\end{align}
The running of the spectral index measured by the Planck team is about $d{n_s}/d\ln k =  - {\rm{0}}{\rm{.0085}} \pm {\rm{0}}{\rm{.0076}}$ (68\% CL, Planck 2015 TT,TE,EE+lowP data) \cite{Planck2015}.

In the next step, we need to obtain $\dot{\phi}$, $\ddot{\phi}$, and $\dddot{\phi}$ in terms of $t$. To this end, using Eqs. (\ref{Hubble}) and (\ref{A2}), we first rewrite Eq. (\ref{Field:SR}) in the following form:
\begin{equation}\label{Field:Int}
\dot{\phi}^{2}\left(1+\frac{\gamma}{M^{4n-1}}\,t^{\lambda-1}\,\dot{\phi}^{2n-1}\right)+ 2 \,\beta \,t^{\lambda-2} =0,
\end{equation}
where
\begin{equation}\label{gama}
\gamma\equiv \frac{6  A n \lambda}{2^n}, \hspace{.5cm} \beta \equiv A \lambda (\lambda-1).
\end{equation}
Then, taking the first and second time derivatives of Eq. (\ref{Field:Int}), we get
\begin{equation}\label{phiddot}
 \ddot{\phi} = \frac{(\lambda-1)\dot{\phi}^2 t^{-\lambda}+2 \beta t^{-2}}{(\gamma/M^{4n-1})(2n+1)\dot{\phi}^{2n}+2 \dot{\phi} t^{1-\lambda}},
 \end{equation}
\begin{eqnarray}
  \label{phidddot}
  \dddot{\phi}=\left(\frac{t^{-\lambda}}{(\gamma/M^{4n-1})(2n+1)\dot{\phi}^{2n}+2 \dot{\phi} t^{1-\lambda}}\right)^2\left[2(\lambda-1)\dot{\phi}^2(-\dot{\phi}
  +t\ddot{\phi})\right.~~~~\nonumber\\-\left.
  4\beta\,t^{-2+\lambda}\big((3-\lambda)\dot{\phi}+t\ddot{\phi}\big)
  -(\gamma/M^{4n-1})(2n+1)t^{-3+\lambda}\dot{\phi}^{2n-1}\right.~\nonumber\\\times\left.
  \big((\lambda-1)\,t^{2}\dot{\phi}^2(\lambda \dot{\phi}+2(n-1)t\ddot{\phi})+
4t^{\lambda}\beta(\dot{\phi}+n \,t\ddot{\phi})\big)\right].
\end{eqnarray}
Note that to evaluate the inflationary observables, it is convenient to replace the cosmic time, $t$, by the $e$-fold number, $N$, which describes the amount of inflation. It is defined as
\begin{equation}\label{efold-definition}
N \equiv \ln \left(\frac{a_e}{a}\right),
\end{equation}
where $a_{e}$ is the scale factor at the end of inflation. This definition is equivalent to
\begin{equation}\label{efold}
dN=-H\,dt.
\end{equation}
The number of $e$-folds at the moment of horizon crossing is
around $N_{*}\approx50-60$, before the end of inflation \cite{Dodelson2003, Liddle2003}.
It should be noted that inflation with the intermediate scale factor (\ref{scale}) suffers from the graceful exit problem in which inflation cannot end by slow-roll violation. To overcome this problem, we follow the approach of \cite{Martin2014b}
and introduce an extra parameter $t_e$ which refers to time in which an unspecified reheating process is triggered to stop inflation. In this way, with the help of Eqs. (\ref{Hubble}), we can solve the differential equation (\ref{efold}) for the intermediate scale factor (\ref{scale}) and get
\begin{equation}\label{t}
t=\left(t_{e}^\lambda-\frac{N}{A}\right)^{1/\lambda},
\end{equation}
where we have used the initial condition $N_e \equiv  N(t_e) = 0$ from Eq. (\ref{efold-definition}).

To see the effect of the Galileon term, let us first study in details the regime ${\cal D}\gg 1$ (or $M\rightarrow 0$); then we focus on the model with general values of ${\cal D}$. It should be remembered that , the standard inflation model corresponds to the limit ${\cal D}\ll 1$ ($M \rightarrow \infty$).

\subsection{Galileon intermediate inflation in the regime ${\cal D}\gg 1$ (or $M\rightarrow 0$)}

In the limit ${\cal D}\gg1$ (or $M\rightarrow 0$), Eq. (\ref{Field:Int}) can be solved analytically to find the field velocity $\dot{\phi}$ as follows:
\begin{equation}\label{phi:dot}
\dot{\phi}=\left(\frac{(1-\lambda) 2^{n} M^{4n-1}}{3  n t}\right)^{\frac{1}{2n+1}}.
\end{equation}
Integrating Eq. (\ref{phi:dot}), we find
\begin{equation}\label{phi}
\phi =\frac{2n+1}{2n}\,\left(\frac{(1-\lambda)\,2^{n} M^{4n-1}}{3\,n}\right)^{\frac{1}{2n+1}} t^{\frac{2n}{2n+1}},
\end{equation}
where we have set the constant of integration to be zero, without loss of generality.
Substituting $t$ from Eq. (\ref{phi}) into (\ref{Hubble}), we obtain
\begin{equation}\label{H}
H=A\,\lambda \left[\left(\frac{3 n}{(1-\lambda)M^{4n-1}2^n}\right)\left(\frac{2n}{2n+1}\right)^{2n+1}\right]^\frac{\lambda-1}{2n}
\phi^\frac{(2n+1)(\lambda-1)}{2n}.
\end{equation}
Finally, by replacing Eq. (\ref{H}) into the first Friedmann equation (\ref{FR:SR}), the form of the inflationary potential in the slow-roll approximation reads
\begin{equation}\label{V}
V=V_{0}\,\phi^{-s},
\end{equation}
where
\begin{equation}\label{s}
s\equiv\frac{2n+1}{n}\,(1-\lambda),
\end{equation}
and
\begin{equation}\label{V0}
V_{0}\equiv 3\, (A\,\lambda)^2 \left[\left(\frac{3n}{(1-\lambda)M^{4n-1}2^n}\right)\left(\frac{2n}{2n+1}\right)^{2n+1}\right]^\frac{\lambda-1}{n}.
\end{equation}
We see that in the Galileon scenario with the functional form of $G(\phi, X)$ as Eq. (\ref{G-form}), in the limit ${\cal D}\gg1$; similar to the standard inflationary model \cite{Barrow90,Barr-Lidd1993,Barrow2006}, intermediate inflation arises from an inverse power-law potential. Note that the potential (\ref{V}) satisfies the condition $V_{,\phi}<0$, as expected in the presence of $\dot{\phi}>0$ to have a Galileon model free of ghosts and Laplacian instabilities.

Using Eqs. (\ref{FR:SR}), (\ref{Hubble}), (\ref{A2}), and (\ref{phi:dot}), the scalar power spectrum (\ref{Ps:A}) turns into
\begin{equation}\label{Ps:A:In}
{\cal P}_{s}=\frac{3\,(A\lambda)^3}{32\, \pi^2\,(1-\lambda)}\,{\sqrt{\frac{3\,n}{2}}}\,t^{3\lambda-2}\,.
\end{equation}
Substituting $t$ from Eq. (\ref{t}) into (\ref{Ps:A:In}), the
power spectrum of the scalar perturbations can be rewritten as
\begin{equation}\label{Ps:int}
{\cal P}_s=\frac{3\,(A\lambda)^3}{32\, \pi^{2}\,(1-\lambda)}\sqrt\frac{3n}{2}
\left(t_{e}^\lambda-\frac{N}{A}\right)^\frac{3\lambda-2}{\lambda}\,.
\end{equation}
Also, using Eqs. (\ref{A2}) and (\ref{epsilon:eta}), with the help of Eq. (\ref{phi:dot}), the scalar spectral index $n_s$ (\ref{ns:A}) and the tensor-to-scalar ratio $r$ (\ref{r:A}) can be obtained in terms of $t$ and the model parameters
\begin{align}
\label{ns:int1}
 n_{s}=1 - \left(\frac{2-3\lambda}{A\,\lambda}\right)t^{-\lambda},\\
 \label{r:int1}
  r= \frac{64}{3}\,\sqrt{\frac{2}{3\,n}}\,\left(\frac{1-\lambda}{\,A\,\lambda}\right)t^{-\lambda}.
 \end{align}
 Applying the approximation $d\ln k \thickapprox H dt$ and Eq. (\ref{Hubble}), from  Eq. (\ref{ns:int1}) we can find the running of the scalar spectral index $d n_{s}/d\ln k$ as
 \begin{equation}\label{dns:int1}
 \frac{d n_{s}}{d\ln k}=\left(\frac{2-3\lambda}{A^2\,\lambda}\right)t^{-2 \lambda}.
 \end{equation}
Now, by replacing Eq. (\ref{t}) in Eqs. (\ref{ns:int1})-(\ref{dns:int1}), we get
\begin{align}
\label{ns:int2}
  n_{s}=1-\frac{2-3\lambda}{A\,\lambda\,(t_{e}^\lambda-\frac{N}{A})}, \\
  \label{r:int2}
  r=\frac{64\,\sqrt{\frac{2}{3n}}\,(1-\lambda)}{3\,A\,\lambda\,(t_{e}^\lambda-\frac{N}{A})}, \\
  \label{dns:int2}
 \frac{d n_{s}}{d\ln k}=\frac{2-3\lambda}{A^2\lambda\,(t_{e}^\lambda-\frac{N}{A})^2}.
\end{align}
From Eq. (\ref{ns:int2}), we find that the parameter $\lambda$ should be in the range of $0<\lambda<2/3$, because for $2/3<\lambda<1$, the blue-tilted spectrum of curvature perturbations $(n_s>1)$ is produced, which is at odds with the prediction from the Planck 2015 data \cite{Planck2015}. The case $\lambda=2/3$ corresponds to the scale-invariant or Harrison-Zel'dovich spectrum $(n_s=1)$, which is also inconsistent with the Planck 2015 observations \cite{Planck2015}.

Now, by fixing Eq. (\ref{Ps:int}) at horizon crossing $e$-fold number $N_*$ as ${\cal P}_{s*}\equiv{\cal P}_s \big|_{N=N_*}=2.207\times10^{-9}$ (68\% CL, Planck 2015 TT,TE,EE+lowP data) \cite{Planck2015}, we find an analytical expression for $t_e$ in terms of $N_*$ and the other model parameters as
\begin{equation}\label{te}
t_{e}=\left[\left(\frac{32\,\pi^2(1-\lambda)}{3\,(A\,\lambda)^3}\,\sqrt{\frac{2}{3 n}}\,{\cal P}_{s*}\right)^\frac{\lambda}{3\lambda-2}+\frac{N_{*}}{A}\right]^\frac{1}{\lambda}.
\end{equation}
Finally, substituting Eq. (\ref{te}) into Eqs. (\ref{ns:int2})-(\ref{dns:int2}), one can rewrite $n_s$, $r$ and $d{n_s}/d\ln k$ in the following forms:
\begin{eqnarray}\label{nsf}
 n_{s}=1-(2-3\lambda)\left(\frac{32 \pi^2(1-\lambda)}{3\,(A\,\lambda)^{2/\lambda}}\,\sqrt{\frac{2}{3 n}}\,{\cal P}_{s*}\right)^\frac{\lambda}{2-3\lambda},
 \end{eqnarray}
\begin{eqnarray}\label{rf}
 r=\frac{64}{3}\,(1-\lambda)\,\sqrt\frac{2}{3\,n}\left(\frac{32\,\pi^2(1-\lambda)}{3\,(A\,\lambda)^{2/\lambda}}\,\sqrt{\frac{2}{3n}}
 \,{\cal P}_{s*}\right)^\frac{\lambda}{2-3\lambda},
 \end{eqnarray}
\begin{eqnarray}\label{dnsf}
 \frac{d n_{s}}{d\ln k}=\frac{2-3\lambda}{A^2 \lambda}\left(\frac{32\,\pi^2(1-\lambda)}{3\,(A\,\lambda)^{3}}\,\sqrt{\frac{2}{3 n}}
 \,{\cal P}_{s*}\right)^\frac{2\lambda}{2-3\lambda}.
 \end{eqnarray}
 As we see, in the limit ${\cal D}\gg 1$, the inflationary observables $n_{s}$, $r$, and $d n_{s}/d\ln k$ are independent of the mass parameter $M$ and the $e$-fold number $N_*$. Therefore, we can combine Eqs. (\ref{nsf}) and (\ref{rf}) to reach the following linear relation
 \begin{equation}\label{r:lin}
r=\frac{64}{3}\,\sqrt{\frac{2}{3n}}\left(\frac{1-\lambda}{2-3\lambda}\right)(1-n_{s}).
\end{equation}
Note that  to satisfy the Planck constraints
on different non-Gaussianity parameters, from Eqs. (\ref{fNL:A}), (\ref{fNLortho:A}) and (\ref{fNLenfold:A}), the range of parameter $n$ should be limited  as $n\leq 212$.
 Using Eq. (\ref{cs2:A}), we also obtain  $ c_{s} \gtrsim0.06$.
  \begin{figure}[t]
\begin{center}
\scalebox{0.9}[0.9]{\includegraphics{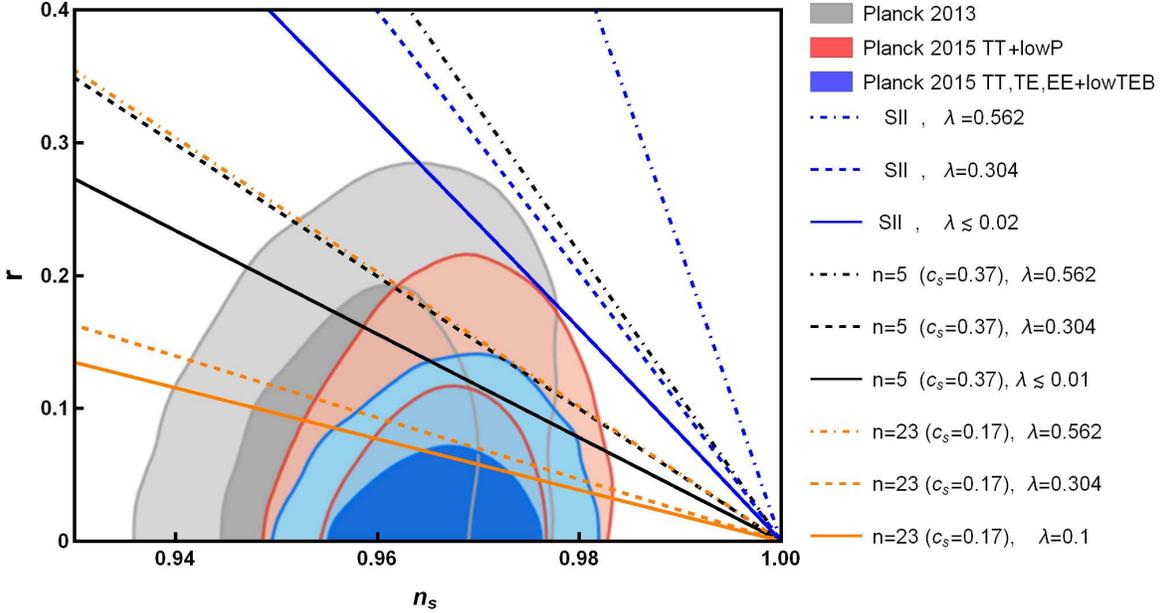}}
\caption{The $r-n_s$ diagram for the intermediate inflation (\ref{scale}) in the Galileon framework (\ref{Lagrangian}) with $K(\phi, X)=X-V(\phi)$ and the Galileon term $G(X)\,\Box \phi=\frac{{X^n}}{{M^{4n-1}}}\,\Box \phi$ in the limit ${\cal D}\gg1$ (or $M \rightarrow 0$) for some typical values of $n$ (or $c_s$) and $\lambda$ with varying $A$ in the range of $A > 0$. Also, the results for the standard intermediate inflation (SII) with different values of $\lambda$ are presented in the figure for comparison. The marginalized joint 68\% and 95\% CL regions of Planck 2013, Planck 2015 TT+lowP and Planck 2015 TT,TE,EE+lowP data \cite{Planck2015} are specified by gray, red, and blue, respectively.}
\label{fig1}
\end{center}
\end{figure}
Now from Eqs. (\ref{nsf}), (\ref{rf}) and using the constraint $n\leq 212$, we plot the $r-n_s$ diagram for the intermediate inflation (\ref{scale}) in the framework of Galileon scalar field in the limit ${\cal D}\gg1$ and compare the prediction of the model with the Planck 2015 observations. This diagram is shown in Fig. \ref{fig1} for some typical values of $n$ and $\lambda$ with varying $A$ in the range of $A > 0$.
 \begin{figure}
  \centering
  \begin{tabular}{ccc}
    \scalebox{0.8}[0.8]{\includegraphics{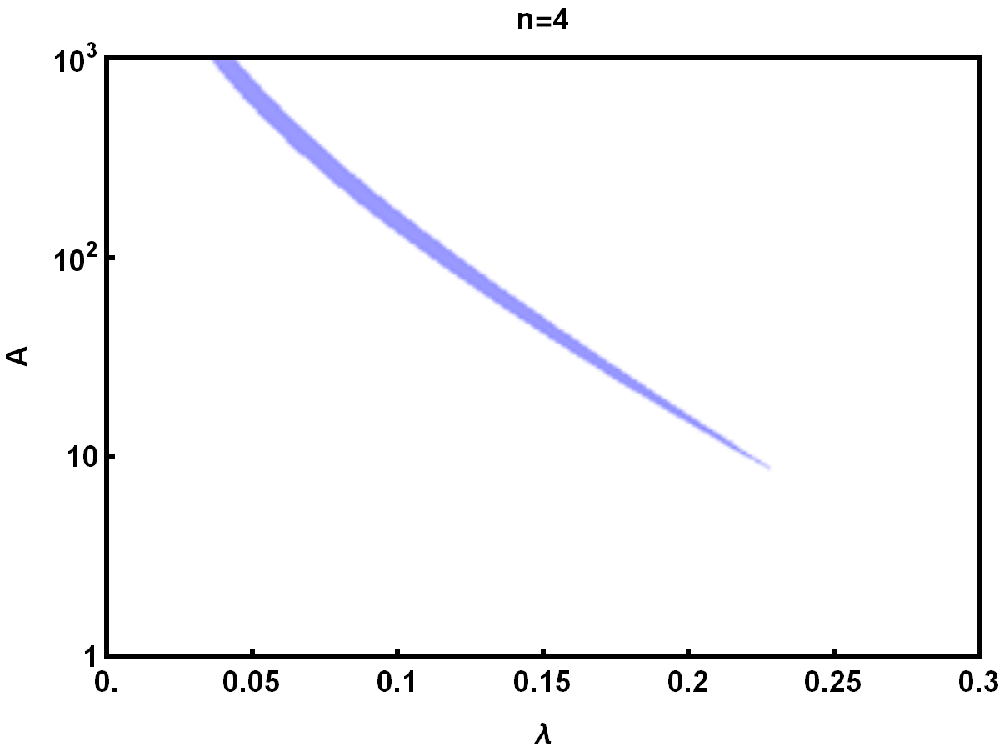}} && \scalebox{0.8}[0.8]{\includegraphics{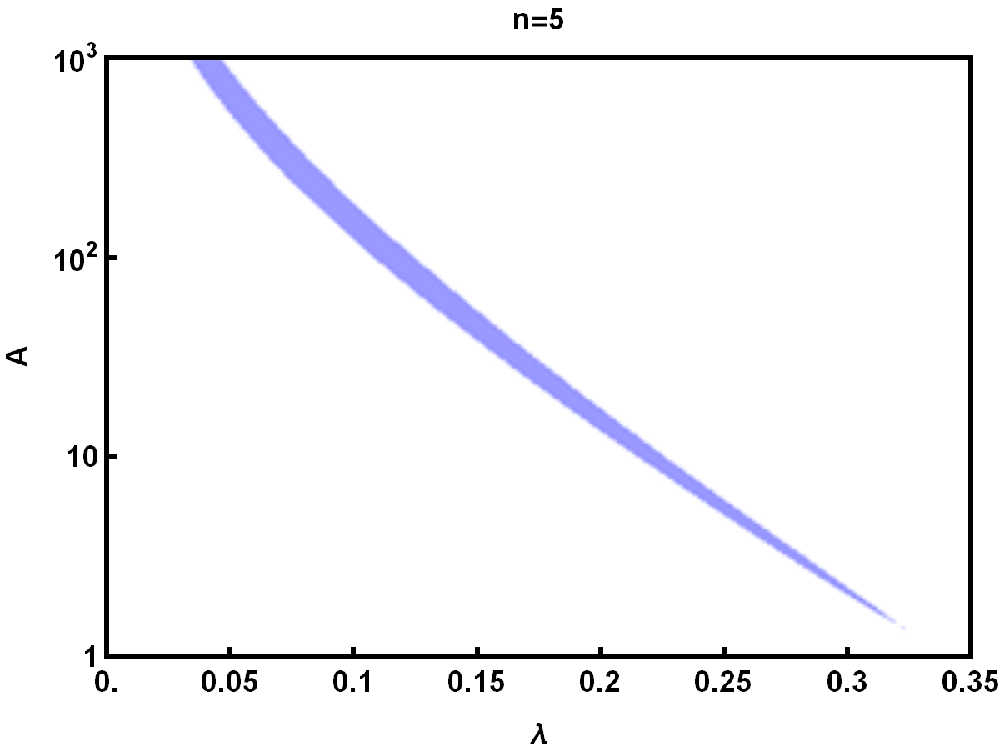}}\\
    \scalebox{0.8}[0.8]{\includegraphics{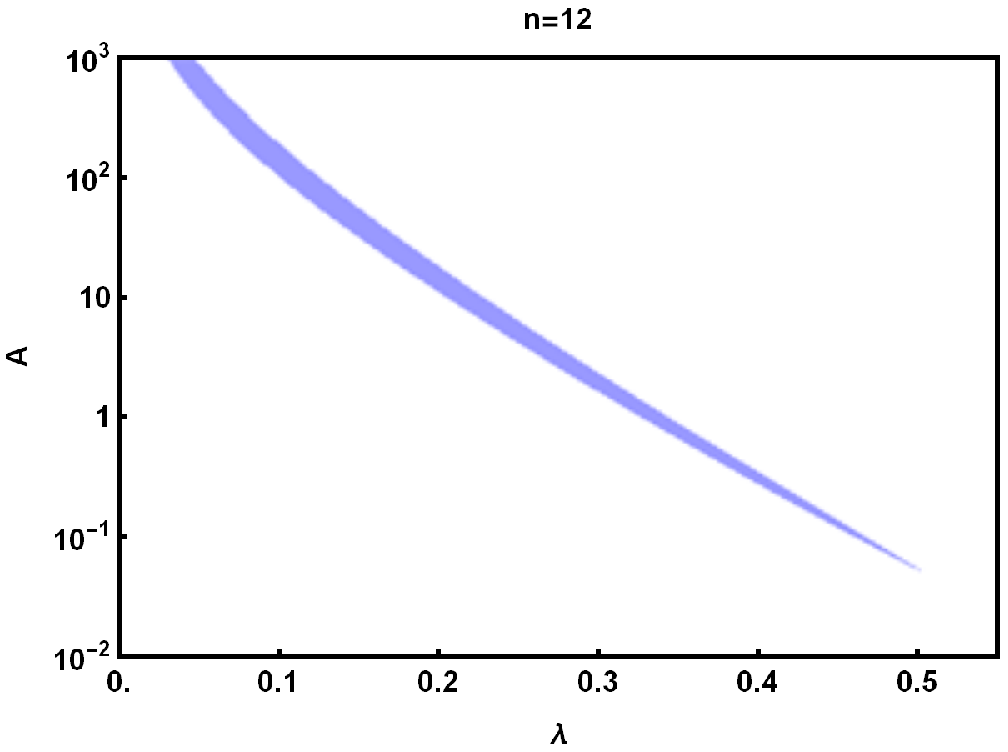}} && \scalebox{0.8}[0.8]{\includegraphics{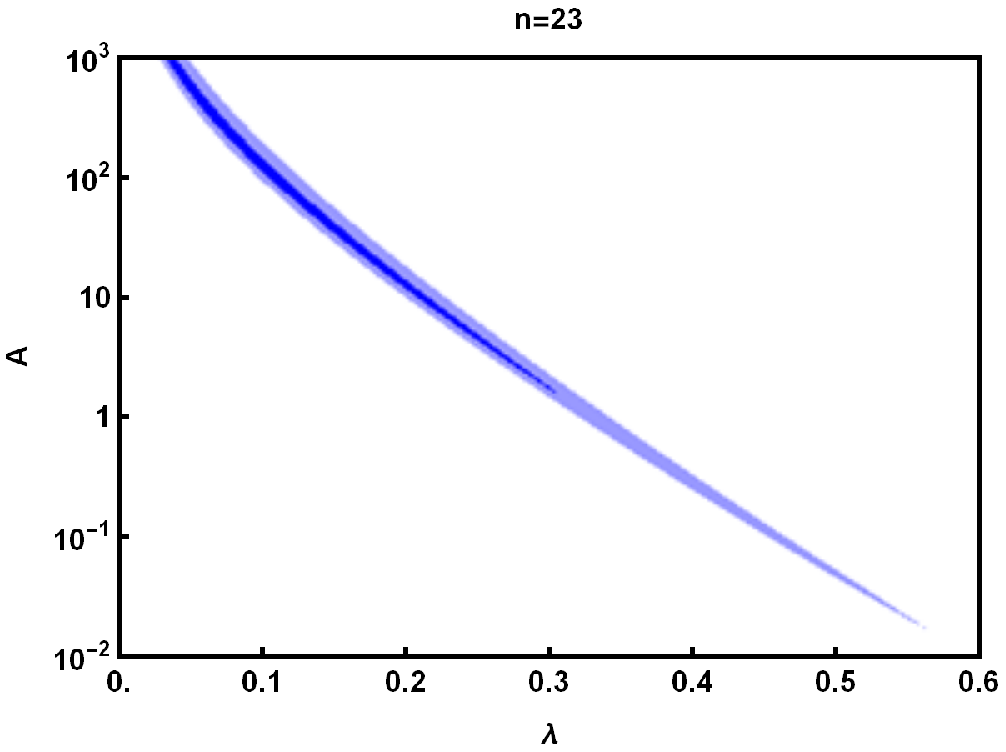}}\\
      \scalebox{0.8}[0.8]{\includegraphics{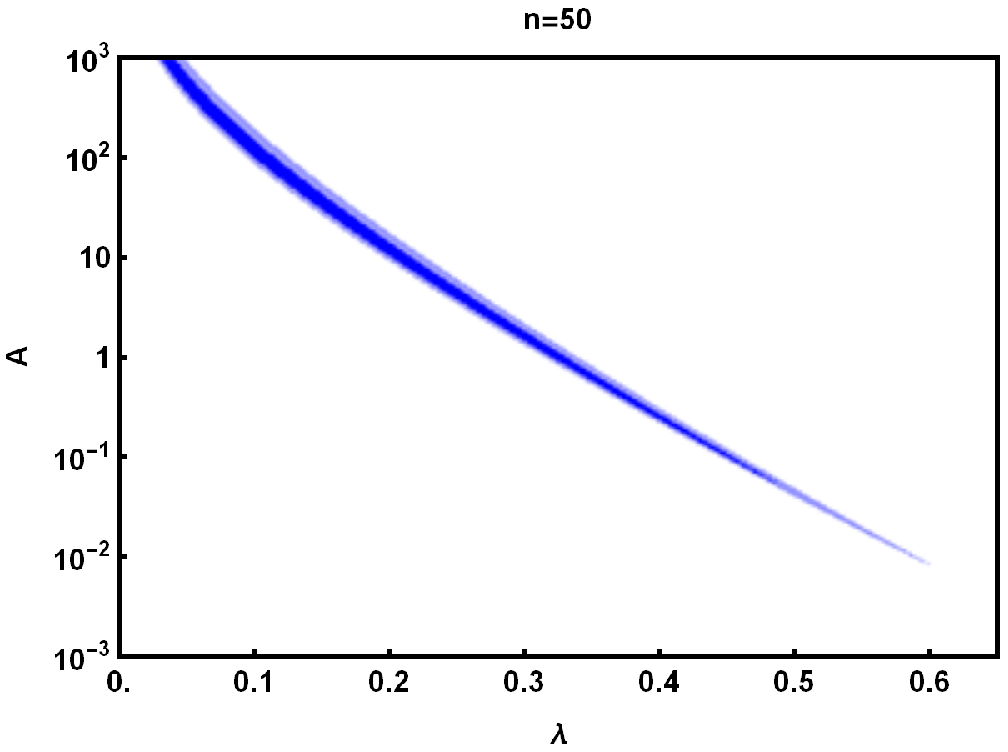}} && \scalebox{0.8}[0.8]{\includegraphics{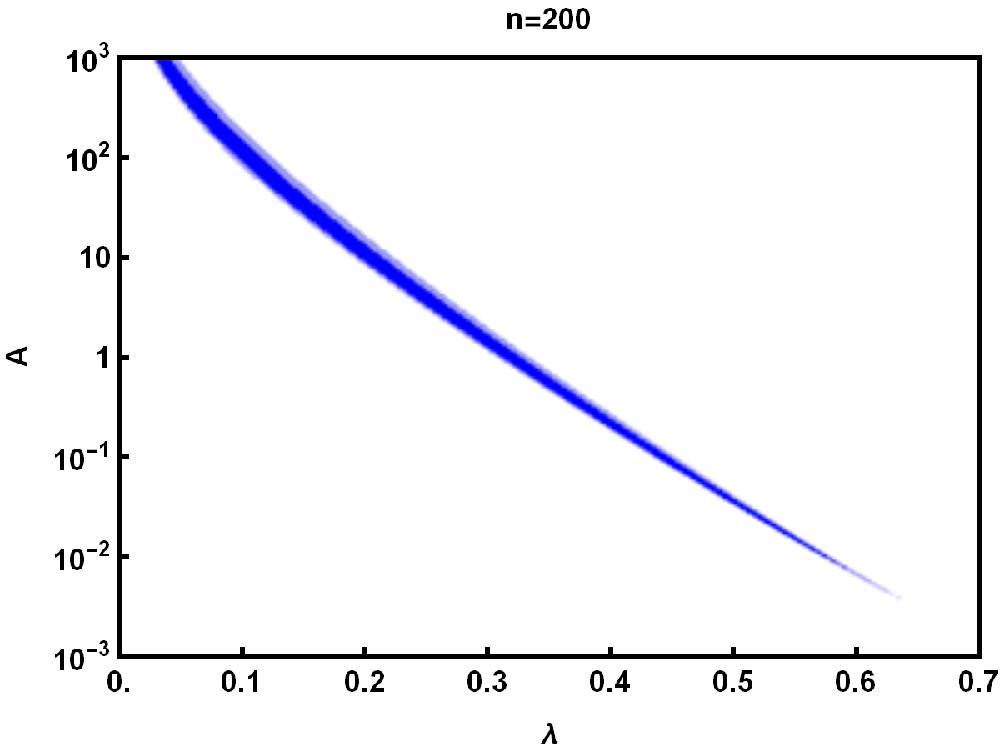}}
  \end{tabular}
\caption {Parameter space of $A$ and $\lambda$ for which the intermediate inflation (\ref{scale}) in the Galileon setting (\ref{Lagrangian}) described by $K(\phi, X)=X-V(\phi)$ and the Galileon term $G(X)\,\Box \phi=\frac{{X^n}}{{M^{4n-1}}}\,\Box \phi$ with different values of $n$ in the limit ${\cal D}\gg1$ (or $M\rightarrow 0$) is compatible with the Planck 2015 data. The darker and lighter blue areas specify the parameter space for which our model is consistent with the Planck 2015 TT,TE,EE+lowP data \cite{Planck2015} at 68\%  and 95\% CL, respectively.}
    \label{fig2}
\end{figure}
  Also, in this figure we represent the prediction of intermediate inflation in the standard setting (see \cite{Barrow2006,Rezazadeh2015,Teimoori2017} for more details). The model of standard intermediate inflation (SII) corresponds to the limit ${\cal D}\ll 1$ $(M \rightarrow \infty)$ in our model. The dashed, dash-dotted, and solid blue lines are corresponding to $\lambda= 0.304$, $0.562$, and $\lambda \lesssim 0.02$, respectively, with varying $A$ in the SII model. Figure \ref{fig1} clears that although the prediction of SII lies completely outside the region allowed by the Planck 2015 data, the result of intermediate inflation in the framework of Galileon scalar field in the limit ${\cal D}\gg1$ can be compatible with the observations.
For instance, from Fig. \ref{fig1}, we see that the results of Galileon intermediate model for
$n=5$ and $n = 23$, respectively, can lie within the 95\% CL and 68\% CL regions allowed by Planck 2015 TT,TE,EE+lowP data \cite{Planck2015}.
As shown in Fig. \ref{fig1}, in the case $n=23$, if $\lambda\lesssim 0.304$ $(\lambda\lesssim 0.562)$, the results of the model are consistent with the 68\% CL (95\% CL) constraint of the Planck 2015 data.

In Table \ref{table1}, using the Planck observational constraints on $r-n_s$ plane and with the help of Eqs. (\ref{nsf}) and (\ref{rf}), we present the ranges of parameter $A$ for which the model for $n= 23$ and some typical values of $\lambda$, in the limit ${\cal D}\gg1$ is in agreement with the Planck 2015 data. In addition, we evaluate the running of the scalar spectral index $d n_{s}/d\ln k$ with the help of Eq. (\ref{dnsf}). Table \ref{table1} shows that the values of $d n_{s}/d\ln k$ predicted by our model are consistent with the $95\%$ CL constraint of Planck 2015 TT,TE,EE+lowP data \cite{Planck2015}.
\begin{table*}[ht!]
  \centering
  \caption{The ranges of parameter $A$, for which the $r-n_s$ diagram of the Galileon intermediate inflation in the limit ${\cal D}\gg1$ with $n=23$ and some typical values of $\lambda$ is compatible with the 68\% or 95\% CL regions of Planck 2015 TT,TE,EE+lowP data \citep{Planck2015}. Furthermore, the estimated values for the running of the scalar spectral index $dn_s/d\ln k$ are presented in the table that satisfy the 95\% CL constraint provided by the Planck 2015 TT,TE,EE+lowP data \cite{Planck2015}.}
\scalebox{1}[1] {\begin{tabular}{c c c c c}
    \hline
    \hline
    $ \quad \lambda \quad $ & $A$ (68\% CL) & \makecell[c]{$\frac{dn_{s}}{d\ln k}$}  & $A$ (95\% CL)& $\frac{dn_{s}}{d\ln k}$\\
    \hline
    $0.1$ & $[112.9,144.9]$ &\,\,\, $[4.190 \times 10^{-5} , 7.537 \times 10^{-5}]$&$[94.0,195.2]$ &\,\,\,$[2.078 \times 10^{-5} , 1.160 \times 10^{-4}]$ \\
    $0.2$ & $[11.8 ,13.9]$  &\,\,\, $[1.086 \times 10^{-4},  1.734 \times 10^{-4}]$&$[10.1 ,18.0]$ &\,\,\,$[5.187 \times 10^{-5},  2.704 \times 10^{-4}]$\\
    $0.35$&      ---        &        ---         & $[0.59 ,0.85]$ & \,\,\, $[1.400 \times 10^{-4},6.513 \times 10^{-4}]$ \\
    $0.4$ &      ---        &       ---         &$[0.246, 0.328]$ &\,\,\,$[1.954 \times 10^{-4} , 8.233 \times 10^{-4}]$ \\
    $0.5$ &      ---        &       ---        & $[0.0456,0.0529]$&\,\,\, $ [4.082\times 10^{-4}  ,1.339\times 10^{-3}]$ \\
    $0.55$&      ---        &       ---        &$[0.0206, 0.0215]$&\,\,\, $  [8.990  \times 10^{-4} , 1.466 \times 10^{-3}]$ \\
    $\lambda \gtrsim 0.562$ & --- & ---   & --- &---\\
    \hline
    \end{tabular}}
  \label{table1}
\end{table*}

It is useful to find the parameter space of $A$ and $\lambda$ for which our model in the limit
${\cal D}\gg1$ is consistent with the Planck 2015 data. This parameter space is illustrated in Fig. \ref{fig2}, for  some specified values of $n$.
The darker and lighter blue areas specify the parameter space for which our model is consistent with the Planck 2015 TT,TE,EE+lowP data \cite{Planck2015} at 68\%  and 95\% CL, respectively. To draw this diagram, we have employed a computational
code that calculates $n_{s}$ and $r$ by using Eqs. (\ref{nsf}) and (\ref{rf}), respectively, and then projects the values of $A$ and $\lambda$ which are consistent with the 68\% or 95\% CL constraints of the Planck 2015 data in
a $2$D contour plot. Note that the constraint on parameter $\lambda$ predicted by our model takes place more or less in the same range obtained for different inflationary frameworks, like the standard canonical inflation \cite{Barrow2006,Campo14}, non-canonical scalar field model \cite{Rezazadeh2015}, $f(T)$-gravity \cite{Rezazadeh2016,Oikonomou2017}, and DBI scenario \cite{Amani2018}.

Furthermore, in Table \ref{table2}, by using Eqs. (\ref{nsf}), (\ref{dnsf}), (\ref{rf}), (\ref{local}), (\ref{fNL:A}), (\ref{fNLortho:A}), and (\ref{fNLenfold:A}), we have estimated the inflationary observables $n_s$, $dn_{s}/{d\ln k}$, $r$, $f_{\rm NL}^{\rm local}$, $f_{\rm NL}^{\rm equil}$, $f_{\rm NL}^{\rm ortho}$, and $f_{\rm NL}^{\rm enfold}$, respectively, in the Galileon intermediate inflation with $\lambda=0.1$ and some typical values of $n$ and $A$. We have also presented the $r-n_{s}$ consistency for each case in the last column of Table \ref{table2}, which is in agreement with that illustrated in Fig. \ref{fig2}. The results are presented in the columns 6-9 and imply that the shape of non-Gaussianities is close to the equilateral one, and is in agreement with  \cite{Felice2011}.
 \begin{table*}[ht!]
  \centering
  \caption{Estimated values of the inflationary observables for $\lambda=0.1$, and some typical values of $n$ and $A$ in the Galileon intermediate model with the Galileon term $G(X)\,\Box \phi=\frac{{X^n}}{{M^{4n-1}}}\,\Box \phi$  in the limit ${\cal D}\gg1$.}
\scalebox{1}[1] {\begin{tabular}{c c c c c c c c c c}
    \hline
    \hline
    $ \quad n \quad $ & $A$ & \,\,\,\,\,\,\,\,\,\,\,\,\makecell[c]{$n_s$}  & $\frac{dn_{s}}{d\ln k}$ &  $r$ & $f_{\rm NL}^{\rm local}$ &$f_{\rm NL}^{\rm equil}$ &$f_{\rm NL}^{\rm ortho}$&$f_{\rm NL}^{\rm enfold}$ & \makecell[c]{$r-n_s$ \\ consistency} \\
    \hline
    $3$ & \,\,\,$150$ &\,\,\, $0.9728$ &\,\,\,$4.354 \times 10^{-5}$& \,\,\,$0.1449$ &\,\,\,$0.011$&\,\,\,$-0.384$& \,\,\, $-0.249$& \,\,\,$-0.068$&\,\,\, --- \\
    $4$ &\,\,\,$150$ &\,\,\,$0.9730$& \,\,\,$4.281 \times 10^{-5}$ &\,\,\,$0.1244$ &\,\,\,$0.011$& \,\,\,$-0.607$&\,\,\,$-0.399$ &\,\,\,$-0.104$&\,\,\,95\% CL\\
    $5$ &\,\,\,$140$&\,\,\,$0.9709$ & \,\,\,$4.970\times 10^{-5}$ &\,\,\, $0.1199$&\,\,\,$0.012$&\,\,\,$-0.829$ &\,\,\, $-0.548$&\,\,\, $-0.141$&\,\,\,95\% CL\\
    $12$ & \,\,\,$125$& \,\,\,$0.9676$ &\,\,\,$6.163\times 10^{-5}$ &\,\,\,$0.0862$&\,\,\,$0.014$&\,\,\,$-2.387$ &\,\,\, $-1.595$&\,\,\, $-0.396$ &\,\,\,95\% CL\\
    $23$ &\,\,\,$125$&\,\,\, $0.9682$&\,\,\,$5.932\times 10^{-5}$&\,\,\, $ 0.0611$&\,\,\,$0.013$&\,\,\,$-4.834$  &\,\,\, $-3.240$&\,\,\, $-0.797$&\,\,\,68\% CL\\
    $50$&\,\,\, $120$& \,\,\,$0.9674$ &\,\,\,$6.238\times 10^{-5}$&\,\,\, $ 0.0425$ &\,\,\,$0.014$&\,\,\,$-10.841$&\,\,\, $-7.278$&\,\,\, $-1.781$ &\,\,\,68\% CL\\
    $200$ &\,\,\,$100$&\,\,\,$0.9613$ &\,\,\,$8.830\times 10^{-5}$ &\,\,\,$0.0253$&\,\,\,$0.016$&\,\,\,$-44.213$& \,\,\,$\,\,\,-29.713$&\,\,\, $-7.250$&\,\,\,68\% CL\\
    \hline
    \end{tabular}}
  \label{table2}
\end{table*}

\subsection{Galileon intermediate inflation in the regime with general $M$}

Now, we are interested to investigate the intermediate inflation in the Galileon scenario (\ref{Lagrangian}) described by Eqs. (\ref{K-form}) and (\ref{G-form}), in the case that covers the general values of the mass scale parameter, $M$. Although the approach is same as that was followed in the limit $M\rightarrow 0$ (or ${\cal D}\gg 1$), for general values of $M$ we need to evaluate the inflationary observables $n_{s}$, $r$, $dn_{s}/d\ln k$, $f_{\rm NL}^{\rm local}$, $f_{\rm NL}^{\rm equil}$, $f_{\rm NL}^{\rm ortho}$  and $f_{\rm NL}^{\rm efold}$, numerically.

Substituting Eq. (\ref{A2}) into Eqs. (\ref{fNL2}), (\ref{fNL-ortho2}), (\ref{fNL-enfold2}), (\ref{Ps-t}), (\ref{ns-t}), (\ref{r-t}) and (\ref{dns-t}), we find the relations
\begin{eqnarray}\label{observables1}
{\cal P}_s={\cal P}_s \big(t,\dot{\phi},n,M,A,\lambda \big),\nonumber\\
n_s=n_s\big(t,\dot{\phi},\ddot{\phi},n,M,A,\lambda\big),\nonumber\\
r=r\big(t,\dot{\phi},n,M,A,\lambda\big),\nonumber\\
\frac{dn_s}{d\ln k}=\frac{dn_s}{d\ln k}\big(t,\dot{\phi},\ddot{\phi},\dddot{\phi},n,M,A,\lambda\big),\nonumber\\
f_{\rm NL}^{\rm equil}=f_{\rm NL}^{\rm equil}\big(t,\dot{\phi},n,M,A,\lambda\big),\nonumber\\
f_{\rm NL}^{\rm ortho}=f_{\rm NL}^{\rm ortho}\big(t,\dot{\phi},n,M,A,\lambda\big),\nonumber\\
f_{\rm NL}^{\rm enfold}=f_{\rm NL}^{\rm enfold}\big(t,\dot{\phi},n,M,A,\lambda\big).
\end{eqnarray}
Then, substituting $n_s=n_s(t,\dot{\phi},\ddot{\phi},n,M,A,\lambda)$ from Eq. (\ref{observables1}) in Eq. (\ref{local}), we can also find
\begin{equation}\label{obser-local}
f_{\rm NL}^{\rm local}=f_{\rm NL}^{\rm local}\big(t,\dot{\phi},\ddot{\phi},n,M,A,\lambda\big).
\end{equation}

Now, for a given set of model parameters $(n,M,A,\lambda\big)$, we solve Eq. (\ref{Field:Int}) numerically to find $\dot{\phi}=\dot{\phi}(t,n,M,A,\lambda\big)$, and consequently from Eqs. (\ref{phiddot}) and (\ref{phidddot}), we obtain $\ddot{\phi}=\ddot{\phi}(t,n,M,A,\lambda\big)$ and $\dddot{\phi}=\dddot{\phi}(t,n,M,A,\lambda\big)$, respectively.
Our numerical results show that the condition $\dot{\phi}>0$ is satisfied in our model, which is necessary to avoid ghosts and Laplacian instabilities.

Using $\dot{\phi}$, $\ddot{\phi}$, and $\dddot{\phi}$ in terms of $t$ and the model parameters $(n,M,A,\lambda\big)$ in Eqs. (\ref{observables1}) and (\ref{obser-local}), we obtain the inflationary observables as
\begin{equation}\label{observables2}
O=O(t,n,M,A,\lambda),
\end{equation}
where $O$ stands for ${\cal P}_s$, $n_s$, $r$, $\frac{dn_s}{d\ln k}$, $f_{\rm NL}^{\rm local}$, $f_{\rm NL}^{\rm equil}$, $f_{\rm NL}^{\rm ortho}$ and $f_{\rm NL}^{\rm enfold}$. Now, by replacing $t$ from Eq. (\ref{t}) into Eq. (\ref{observables2}), it reads
\begin{equation}
O=O(t_e,N,n,M,A,\lambda).
\end{equation}
To find the parameter $t_e$ in terms
of the other model parameters, we fix the amplitude of
scalar perturbations ${\cal P}_s={\cal P}_s\big(t_{e},N,n,M,A,\lambda\big)$ at the epoch of horizon crossing with the $e$-fold number
$N_{*}$ as  ${\cal P}_{s*}\equiv{\cal P}_s \big|_{N=N_*}=2.207\times10^{-9}$ (68\% CL, Planck 2015 TT,TE,EE+lowP data) \cite{Planck2015}. Interestingly enough,  our numerical analysis implies that like the case $M\rightarrow 0$, for the general values of $M$, the inflationary observables are independent of the $e$-fold number $N_*$ at all. Therefore, the inflationary observables $n_s$, $r$, $\frac{dn_s}{d\ln k}$, $f_{\rm NL}^{\rm local}$, $f_{\rm NL}^{\rm equil}$, $f_{\rm NL}^{\rm ortho}$, and $f_{\rm NL}^{\rm enfold}$ in our Galileon intermediate model depend on the four free parameters $n$, $M$, $A$ and $\lambda$.
\begin{figure}[t]
\begin{center}
\scalebox{0.9}[0.9]{\includegraphics{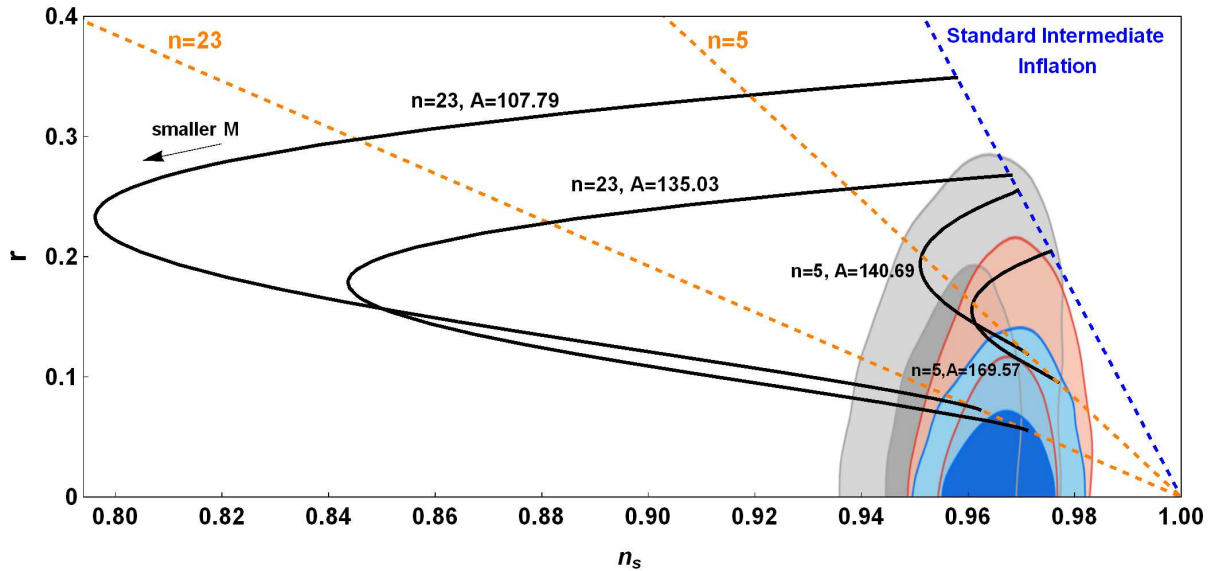}}
\caption{The $r-n_s$ diagram for the intermediate inflation (\ref{scale}) in the Galileon framework with $K(\phi, X)=X-V(\phi)$ and the Galileon term $G(X)\,\Box \phi=\frac{{X^n}}{{M^{4n-1}}}\,\Box \phi$ for some typical values of $n$ and $A$ with $\lambda=0.1$ and varying $M>0$. Also, the results of the Galileon intermediate model in the limit $M\rightarrow 0$ (or ${\cal D}\gg1$), for two different values of  $n=5, 23$  with varying $A>0$ and $\lambda=0.1$ are presented by the orange dashed lines. Furthermore, the prediction of standard intermediate inflation ($M\rightarrow\infty$ or ${\cal D}\ll 1$), with $\lambda=0.1$ is shown by the blue dashed line in the figure for comparison. The marginalized joint 68\% and 95\% CL regions of Planck 2013, Planck 2015 TT+lowP and Planck 2015 TT,TE,EE+lowP data \cite{Planck2015} are specified by gray, red and blue, respectively.}
\label{fig3}
\end{center}
\end{figure}

Figure \ref{fig3} shows the predictions of our Galileon intermediate model in the $r-n_{s}$ plane for some typical values of $n$ and $A$ with $\lambda=0.1$ and varying $M>0$ (see the solid black curves). Also, the  orange dashed lines illustrate the results of model in the limit $M\rightarrow 0$ (or ${\cal D}\gg 1$) for $n = 5$ and 23 with $\lambda=0.1$ and varying $A>0$. Besides, the blue dashed line corresponds to the standard intermediate inflation (SII) with $\lambda=0.1$. Figure \ref{fig3} clears that in the limit $M\rightarrow \infty$ (or ${\cal D} \ll 1$) which corresponds to the standard setting; the intermediate inflation lies outside the 95\% CL region allowed by Planck 2015 TT,TE,EE+lowP data \cite{Planck2015}, while we can make it compatible with the observations in the framework of Galileon scenario. From the figure, we see that as the mass parameter $M$ decreases, the model shows the better consistency with the observations deduced from the Planck 2015 data.
\begin{figure}[t]
\begin{center}
\scalebox{0.9}[0.9]{\includegraphics{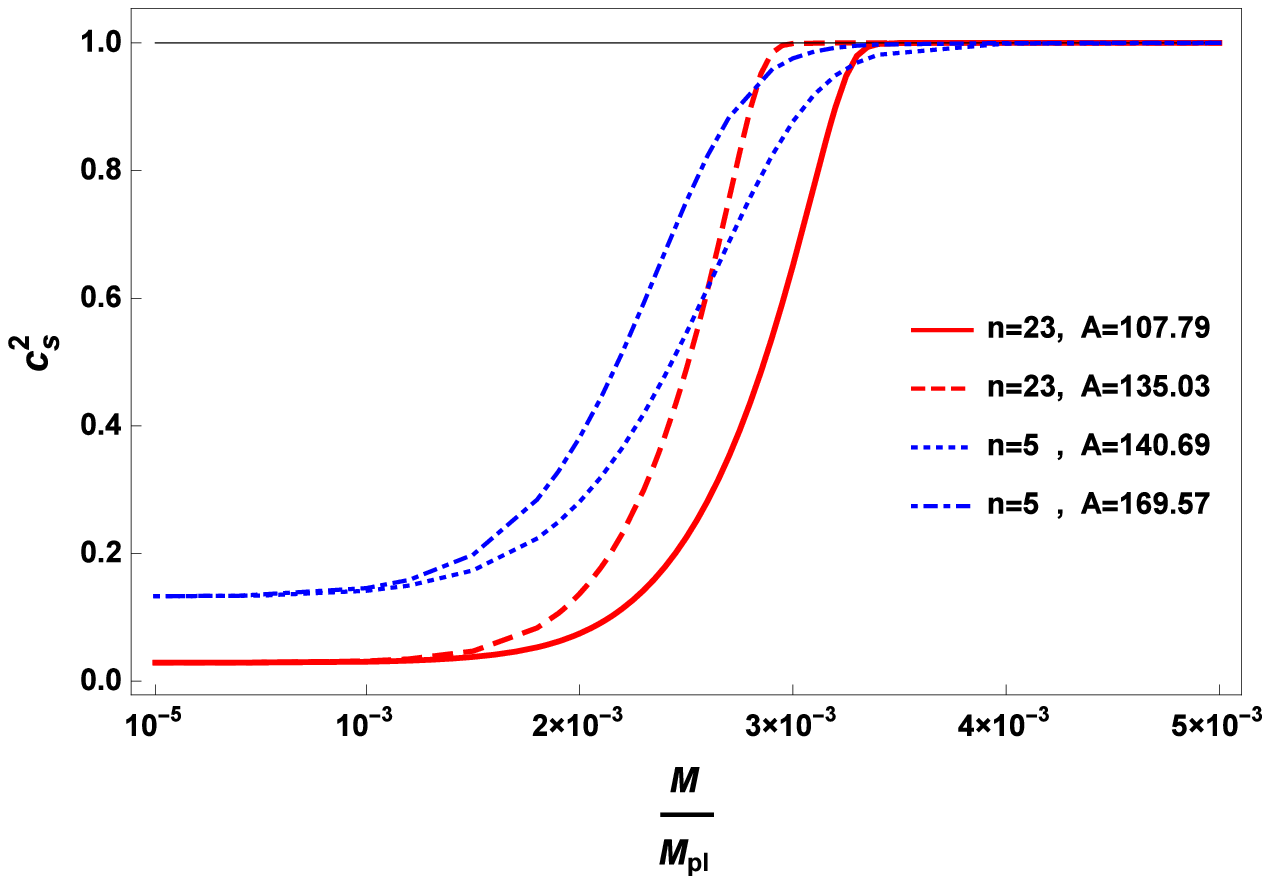}}
\caption{Variation of the scalar propagation speed squared $c_{s}^2$ versus the dimensionless mass parameter $M/M_{\rm pl}$ with $\lambda=0.1$ and some typical values of $n$ and $A$.}
\label{fig4}
\end{center}
\end{figure}
Note that as $M$ decreases, the observables $n_{s}$ and $r$ vary from the values given by the SII, i.e., the regime $M\rightarrow \infty$, and approach to the values given by Eqs. (\ref{nsf}) and (\ref{rf}) valid in the regime $M\rightarrow 0$, as expected.

 In Fig. \ref{fig4}, with the help of Eqs. (\ref{cs2}) and (\ref{A2}), we plot the variation of the scalar propagation speed squared $c_{s}^2$ versus the dimensionless mass parameter $M/M_{\rm pl}$ for some typical values of $n$ and $A$ with $\lambda = 0.1$. Figure \ref{fig4} shows that (i) for small values of $M/M_{\rm pl}$, the sound speed squared $c_s^2$ goes to $2/(3n)$, which is in exact agreement with that obtained in Eq. (\ref{cs2:A}) for the Galileon intermediate inflation in the limit ${\cal D}\gg1$ (or $M\rightarrow 0$); (ii) for large values of $M/M_{\rm pl}$, $c_{s}^2$ approaches 1 for any given values of $n$ and $A$. This is expected, because our Galileon intermediate model in the limit $M\rightarrow \infty$ behaves like the SII characterized by $c_s^2=1$.

\begin{figure}[t]
\begin{center}
\scalebox{0.9}[0.9]{\includegraphics{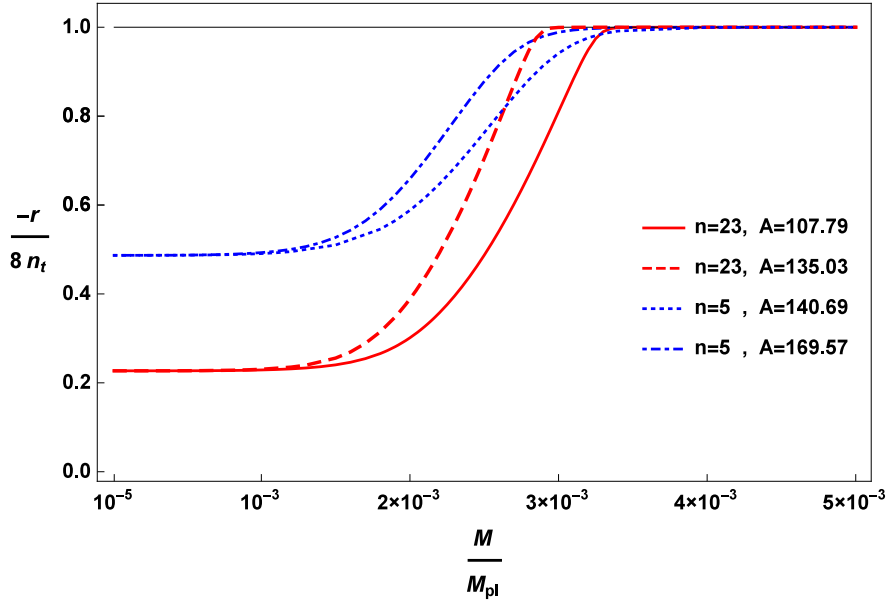}}
\caption{Variation of the ratio $-r/(8n_{t})$ versus the dimensionless mass parameter $M/M_{\rm pl}$ with $\lambda=0.1$ and some typical values of $n$ and $A$.}
\label{fig5}
\end{center}
\end{figure}

In Fig. \ref{fig5}, using Eq. (\ref{cs2}), the consistency relation (\ref{consistency2}) in the Galileon intermediate inflation and Eq. (\ref{A2}), we plot the variation of $-r/(8n_t)$ versus the fractional mass parameter $M/M_{\rm pl}$ for different values of $n$ and $A$ with $\lambda = 0.1$. For comparison, we also plot the consistency relation in the standard scenario characterized by $-r/(8n_t)=1$. Figure \ref{fig5} shows that (i) in the limit of small $M/M_{\rm pl}$, the ratio $-r/(8n_t)$ tends toward $\frac{4}{3}(\frac{2}{3n})^{1/2}$, which is in concurrence with Eqs. (\ref{consistency:A}) and (\ref{cs2:A}) governing the Galileon intermediate inflation in the regime ${\cal D}\gg1$ (or $M\rightarrow 0$);  (ii) in the limit of large $M/M_{\rm pl}$, the ratio $-r/(8n_t)$ approaches the standard consistency relation i.e., $-r/(8n_t)=1$, as expected.

In Table \ref{table3}, with the help of Eqs. (\ref{A2}), (\ref{ns-t}),  and (\ref{r-t}), we have estimated the ranges of the mass scale $M$ for which our model with $\lambda= 0.1$, and some typical values of $n$ and $A$ is consistent with the Planck 2015 data.
For instance, for $n=23$ and $A=135.03$, our Galileon intermediate model can enter the 68\% CL region of the Planck 2015 observations, provided $M\lesssim 15.3\times10^{-4} M_{\rm pl}$.
 In Table \ref{table3}, using Eqs. (\ref{A2}), and (\ref{dns-t}), we also present the predictions of the model for the running of the scalar spectral index $d{n_s}/d\ln k$  which is consistent with the 95\%  CL constraint of the Planck 2015 TT,TE,EE+lowP data \cite{Planck2015}.
Furthermore, with the help of Eqs. (\ref{A2}) and (\ref{ns-t}), from Eqs. (\ref{local}), (\ref{fNL2}), (\ref{fNL-ortho2}), and (\ref{fNL-enfold2}) we estimate the local, equilateral, orthogonal and  enfolded non-Gaussianity parameters, respectively. The results are presented in Table  \ref{table4} and reflect the fact that in our Galileon framework the shape of non-Gaussianity is approximately close to the equilateral one. This result is in agreement with \cite{Felice2011}.
 The estimated values of $f_{\rm NL}^{\rm equil}$  are in well agreement with the  68\%  CL constraint of the Planck 2015 TT,TE,EE+lowP data \cite{Planck2015-nonGauss}.
\begin{table*}[ht!]
  \centering
  \caption{The ranges of parameter $M$, for which the $r-n_s$ diagram of the Galileon intermediate inflation with $\lambda=0.1$ and some typical values of $n$ and $A$ is compatible with the 68\% or 95\% CL regions of Planck 2015 TT,TE,EE+lowP data \citep{Planck2015}.
  Furthermore, the estimated values for the running of the scalar spectral index $dn_s/d\ln k$  and different non-Gaussianity parameters are presented in the table. The results
  show that the shape of non-Gaussianity is approximately close to the equilateral one. Also, the scalar spectral index and equilateral non-Gaussianity values satisfy the 95\% and 68\% CL  constraints of Planck 2015 TT,TE,EE+lowP data \cite{Planck2015,Planck2015-nonGauss}, respectively. This table is continued in TABLE \ref{table4}}
  \begin{tabular}{c c c c c  }
    \hline
    \hline
    $\quad n \quad$ & $  \quad A \quad $ & $M$ &\makecell[c]{$r-n_s$ \\ consistency}  & $\frac{dn_{s}}{d\ln k}$\\
    \hline
    $5$  & $140.69$  &\,\,\,\, $ M\lesssim 18.3 \times 10^{-4} M_{\rm pl}$  & 95\% CL  &\,\,\, $[-2.955 \times 10^{-4} , 4.913 \times 10^{-5}]$ \\
    $5$  & $169.57$  &\,\,\,\, $ M\lesssim 20.0 \times 10^{-4} M_{\rm pl}$  & 95\% CL &\,\,\, $[-3.701 \times 10^{-4}, 3.166 \times 10^{-5}]$ \\
     $23$ & $107.79$  &\,\,\,\, $ M\lesssim 16.6 \times 10^{-4} M_{\rm pl}$  & 95\% CL &\,\,\, $[-5.437 \times 10^{-4} , 8.405 \times 10^{-5}]$ \\
    $23$ & $135.03$  & \,\,\,\,$ M\lesssim 15.3 \times 10^{-4} M_{\rm pl}$  & 68\% CL  &\,\,\, $[-5.464 \times 10^{-4} , 4.947\times 10^{-5}]$\\
    \hline
    \end{tabular}
  \label{table3}
\end{table*}
\begin{table*}[ht!]
  \centering
  \caption{This table is the Continuation of TABLE \ref{table3}.}
  \begin{tabular}{ c c c c }
    \hline
    \hline
    $f_{\rm NL}^{\rm local}$&$f_{\rm NL}^{\rm equil}$ &$f_{\rm NL}^{\rm ortho}$&$f_{\rm NL}^{\rm enfold}$\\
    \hline
   $[0.012,0.015]$  & $[-0.829,-0.658]$  &  $[-0.548,-0.443]$  &  $[-0.141,-0.107]$ \\
   $[0.010,0.015]$  & $[-0.829,-0.414]$  &  $[-0.548,-0.257]$  &  $[-0.141,-0.078]$ \\
   $[0.016,0.019]$  & $[-4.834,-4.200]$  &  $[-3.240,-2.854]$  &  $[-0.797,-0.673]$ \\
   $[0.012,0.015]$  & $[-4.834,-3.978]$  &  $[-3.240,-2.682]$  &  $[-0.797,-0.648]$ \\
    \hline
    \end{tabular}
     \label{table4}
\end{table*}
\section{Conclusions}\label{sec:con}

Here, we investigated intermediate inflation within the framework of Galileon scalar field described by the Lagrangian (\ref{Lagrangian}) with the Galileon term $G(\phi, X)\Box \phi=\frac{X^n}{M^{4n-1}}\,\Box \phi$, where $n>0$ and $M>0$ are integer constant and mass scale parameter, respectively. Using the scalar and tensor power spectrum, we first obtained the fundamental relations governing the inflationary observables including the scalar spectral index $n_{s}$, the tensor-to-scalar ratio
$r$, the running of the scalar spectral index $dn_{s}/d\ln k$, and the non-Gaussianity parameters. We applied these results for the intermediate inflation characterized by the scale factor  $a(t)= a_{0} \exp\left(A\,t^\lambda\right)$, where $A>0$ and $0<\lambda<1$ are constant parameters. Note that the prediction of the intermediate inflation in the standard scenario is completely ruled out by the Planck 2015 data. This motivated us to investigate the intermediate inflation in the Galileon scenario to see whether this model can be resurrected in light of the Planck 2015. We first showed that in the limit $M \rightarrow \infty$, our Galileon intermediate model recovers the results of standard inflation. Then, we turned to study the regime $M \rightarrow 0$ in which the Galileon self-interaction term $G(\phi, X)\Box \phi$ dominates over the standard kinetic term $X$. In this limit, we obtained analytical relations for the inflationary observables and found the following results:
\begin{itemize}
\item The scalar propagation speed, $c_s$, only depends on $n$.

\item The shape of non-Gaussianity is approximated close to the equilateral type. Also, taking into account the Planck 2015 observational constraint on different
     non-Gaussianity parameters, the parameter $n$ is bounded as $n\leq 212$ (or $c_{s}\gtrsim 0.06$).

\item The Galileon intermediate inflation like the standard intermediate model is described by an inverse power-law potential.

\item The predictions of the Galileon intermediate model in the $r-n_{s}$ plane, for some typical values of $n$ and $\lambda$, can lie inside the allowed regions of Planck 2015 TT,TE,EE+lowP data \cite{Planck2015}. For instance, for $n = 23$, if $\lambda \lesssim 0.304$ $(\lambda \lesssim 0.562)$ the results of the model can enter the 68\% CL ($95\%$ CL) region allowed by Planck 2015 data (see Fig. \ref{fig1}).

\item We also determined the parameter space of $A$ and $\lambda$ for some typical values of $n$, for which our model is compatible with the Planck 2015 observations (see Fig. \ref{fig2}).
\end{itemize}

In the regime with general values of $M$, we used a numerical approach to evaluate
the inflationary observables. Our numerical results are summarized as follows:
\begin{itemize}
\item We concluded that as the value of the mass parameter $M$ decreases, the model can be consistent with the observations deduced from the Planck 2015.

\item We found that in the limit of small $M/M_{\rm pl}$, both the scalar propagation speed squared $c_{s}^2$ and the ratio $-r/(8n_t)$, respectively, tend toward $2/(3n)$ and $\frac{4}{3}(\frac{2}{3n})^{1/2}$. Also in the limit of large $M/M_{\rm pl}$, both $c_{s}^2$ and the $-r/(8n_t)$ approach the corresponding standard relations $c_s^2=1$ and $r=-8n_t$, respectively, as expected.

\item We obtained the ranges of $M$ for which the result of the model is in agreement with the 68\%  or 95\% CL constraints of Planck 2015 observations. For instance, we found that the case
 $(n, A)=(23,135.03 )$ is in consistency with the 68\% CL constraint of Planck 2015 data for $M\lesssim 15.3\times10^{-4} M_{\rm pl}$. In addition, we estimated the running of the scalar spectral index $dn_s/d \ln k$ predicted by the model and concluded that it is in well agreement with the 95\%  CL constraint of the Planck 2015 data. We also evaluated the local, equilateral, orthogonal and enfolded non-Gaussianity parameters. Our results showed that the shape of non-Gaussianities is close to the equilateral one and it satisfies the 68\% CL constraint of the Planck 2015 data.
\end{itemize}

\subsection*{Acknowledgements}

The authors thank the anonymous referee for the very valuable comments. The authors
also thank Tsutomu Kobayashi for useful discussions. The work of K. Karami has been supported financially
by Research Institute for Astronomy and Astrophysics of Maragha (RIAAM) under research project No. 1/5440-61.

\end{document}